\documentclass[fleqn,usenatbib]{mnras}

\usepackage[T1]{fontenc}
\usepackage{ae,aecompl}

\usepackage{graphicx}	
\usepackage{amsmath}	
\usepackage{amssymb}	
\usepackage{newtxtext,newtxmath}

\defcitealias{Roberts19}{R19} 

\title[Too dense to go through]{Too dense to go through: The role of low-mass clusters in the pre-processing of satellite galaxies}

\author[D. Pallero et al.]{
Diego Pallero$^{1,2}$\thanks{E-mail: dpallero@dfuls.cl},
Facundo A. G\'omez$^{1,3}$,
Nelson D. Padilla$^{4,5}$,
Yannick M. Bah\'e$^{6}$ ,\newauthor
Cristian A. Vega-Mart\'inez$^{1,3}$,
S. Torres-Flores$^{1}$
\\
$^{1}$Departamento de astronom\'ia, Universidad de La Serena, Avenida Juan Cisternas 1200, La Serena, Chile\\
$^{2}$Instituto de F\'isica y Astronom\'ia, Universidad de Valpara\'iso, Avda. Gran Bretaña 1111, Valpara\'iso, Chile\\
$^{3}$Instituto de Investigaci\'on Multidisciplinar en Ciencia y Tecnolog\'ia, Universidad de La Serena, Ra\'ul Bitr\'an 1305, La Serena, Chile\\
$^{4}$Instituto de Astrof\'isica, Pontificia Universidad Cat\'olica de Chile, Vicuña Mackenna 4860, Santiago, Chile \\
$^{5}$Centro de Astro-Ingenier\'ia, Pontificia Universidad Cat\'olica de Chile, Vicuña Mackenna 4860, Santiago, Chile\\
$^{6}$Leiden Observatory, Leiden University, PO Box 9513, NL-2300 RA Leiden, the Netherlands\\
}
\date{Accepted XXX. Received YYY; in original form ZZZ}

\pubyear{2015}

\begin{document}
\label{firstpage}
\pagerange{\pageref{firstpage}--\pageref{lastpage}}
\maketitle

\begin{abstract}
We study the evolution of  satellite galaxies in clusters of the \textsc{c-eagle} simulations, a suite of 30 high-resolution cosmological hydrodynamical zoom-in simulations based on the \textsc{eagle} code. We find that the majority of galaxies that are quenched at $z=0$ ($\gtrsim$ 80$\%$) reached this state in a dense environment (log$_{10}$M$_{200}$[M$_{\odot}$]$\geq$13.5). 
At low redshift, regardless of the final cluster mass, galaxies appear to reach their quenching state in low mass clusters. Moreover, galaxies quenched inside the cluster that they reside in at $z=0$ are the dominant population in low-mass clusters, while galaxies quenched in a different halo dominate in the most massive clusters. When looking at clusters at $z>0.5$, their in situ quenched population dominates at all cluster masses. This suggests that galaxies are quenched inside the first cluster they fall into.  
After galaxies cross the cluster's $r_{200}$ they rapidly become quenched ($\lesssim$ 1Gyr). Just a small fraction of galaxies ($\lesssim 15\%$) is capable of retaining their gas for a longer period of time, but after 4Gyr, almost all galaxies are quenched.
This phenomenon is related to ram pressure stripping and is produced when the density of the intracluster medium reaches a threshold of $\rho_{\rm ICM}$ $\sim 3 \times 10 ^{-5}$ n$_{\rm H}$ 
(cm$^{-3}$). These results suggest that galaxies start a rapid-quenching phase shortly after their first infall inside $r_{200}$ and that, by the time they reach $r_{500}$, most of them are already quenched.
\end{abstract}

\begin{keywords}
galaxies:clusters: general -- galaxies: evolution -- galaxies: formation -- galaxies: star formation -- galaxies: haloes
\end{keywords}



\section{Introduction}

\defcitealias{Pallero19}{P19}

Much effort has been devoted to understanding the physical processes that drive galaxy evolution. It has been well established that the environment plays a decisive role in shaping important properties of galaxies  \citep[]{Gunn72,Dressler80,Dressler84,Moore96,Poggianti01,Blanton09};
while galaxies located in low-density environments typically show bluer colours and late-type morphology, galaxies located in dense environments, such as galaxy clusters, show redder colours and an early-type morphology  \citep[]{Gomez03, Kauffmann04, Poggianti06,Moran07,Blanton09}. Even within clusters, there are differences between the population of galaxies located on the outskirts and those in the inner, denser regions  \citep[]{Postman05,Fasano15,Haines15,Brough17,Cava17}.

This colour/morphology-density relation can be explained as a decrease in the star formation activity of galaxies located in denser environments and it is thought to be a consequence of the cold gas depletion of the satellite population \citep[]{Haynes84,Gavazzi06,Fabello12,Catinella13,Hess13}. Nevertheless, the main mechanism responsible for this gas depletion is still an open question.
There are several processes that can cause a sharp decrease in the star formation activity of a galaxy. These can be broadly classified as events of  \textit{`mass quenching'} and \textit{`environmental quenching'} \citep{Peng10}. The former corresponds to mechanisms related to internal galaxy processes influenced by the galaxy mass such as gas outflows driven by the presence of an active galaxy nucleus (AGN) \citep[]{Croton_2006,Fabian12,Cicone14}, or supernova feedback and stellar winds \citep[]{Larson74,Dekel86,Efstathiou_2000,Cantalupo_2010}. Environmental quenching, on the other hand, refers to mechanisms related to the interaction between a galaxy and its local environment. Simulations and models have shown that in particular for the case of galaxy groups and clusters, the main mechanisms associated with environmental quenching can be separated into three broad categories \citep
[for details see][]{Boselli06,Jaffe16}: gravitational interactions between galaxies, gravitational interactions between cluster, and galaxies and interactions between galaxies and the intracluster medium (ICM).

It is expected that at least some of these mechanisms act simultaneously, thus rendering the characterization of galaxy quenching in dense environments challenging.
Nevertheless, over the last decade, several studies have suggested that ram pressure and starvation are the main driven responsible for environmental quenching \citep[]{deLucia12,Wetzel13,Muzzin14,Wetzel15,Peng15,Foltz18}.
\citet{Wetzel12} presented a model in which quenching is driven in a \textit{`delayed-then-rapid'} fashion in which galaxies, after its infall into a massive halo, keep forming stars as a central for several Gyr and, once the satellite star formation quenching begins, occurs rapidly.

In a recent observational study, \citet[][hereafter R19]{Roberts19} showed that, inside clusters, galaxies experience what they call a \textit{`slow-then-rapid'} quenching scenario, especially for low-mass galaxies. In the scenario proposed by \citetalias{Roberts19}, when a galaxy first enters the virial radius, $R_{\rm vir}$, of a cluster it starts a slow quenching phase that can last between 1-2.5 Gyr until the galaxy reaches a region where the characteristic density of the ICM reaches $\rho_{\rm ICM} \sim 10^{-28.3}\rm gr~cm^{-3}$ ($\sim 3\times10^{-5}n_{\rm H}~\rm cm^{-3}$). At this point, the galaxy starts a rapid quenching phase, that can take between 0.5-1 Gyr. The authors suggest that this rapid quenching phase could be associated to ram pressure stripping events, but this is hard to conclude from observations alone.

Cosmological simulations are a powerful tool that allows us to disentangle the effects associated with the different processes that are typically coupled in a non-linear fashion. In the last years, analytic models \citep[]{Fujita04,Mok14,Contini20}, semi-analytic \citep[]{deLucia12,Wetzel13,Henriques17,Stevens17,Cora18b,Contini19} and hydrodynamical simulations \citep[e.g.][]{Bahe13,Bahe15,Taylor17,Wright19,Pallero19,Donnari20} have allowed us to explore the relative importance of the different mechanisms that bring a galaxy to a final quenched state. In \citet[hereafter P19]{Pallero19}, we studied the satellite population of galaxy clusters in the publicly available data from the \textsc{eagle} simulation. We found that most of these galaxies reach their quenching state inside galaxy clusters. However, our results suggested that the fraction of galaxies that reach clusters already quenched grows with the final cluster mass.

Quenching of satellites with stellar masses $\gtrsim 10^{9}$ M$_{\odot}$ typically took place in structures with masses in the range of low-mass clusters ($\sim 10^{14}$ M$_{\odot}$).  Unfortunately, our study suffered from low number statistics as \textsc{eagle} only included 10 low and intermediate-mass clusters. 

In this work, we will add new evidence to address these open issues and to disentangle the mechanisms that finally lead to the quenching of the star formation in the satellite population of galaxy clusters.

To this end, we present a detailed study of the quenching history of the satellite population in galaxy clusters from the \textsc{cluster-eagle} simulations \citep[]{Bahe17,Barnes17}, 
a set of zoom-in high-resolution hydrodynamical simulations of 30 galaxy clusters spanning a mass range of 14 $\leq \log_{10} M_{200}^{z=0}$[M$_{\odot}$]$\leq 15.4$ .
Thanks to the wide mass range of the modelled clusters, the same mass resolution as \textsc{eagle} simulations, and an improved time-resolution at low $z$, \textsc{c-eagle}  is an ideal laboratory to study which is the dominant quenching mechanism associated with environmental quenching. This paper is organized as follows. 
Section 2, contains a brief summary of the simulations and a description of the method to identify structures in space and time. In Section 3, we characterize the properties of each cluster and their satellite populations at $z=0$ and study how their properties evolve as a function of the time they spent inside clusters. In Section 4, we discuss how, in our models, ram pressure can account almost completely for the quenching of satellite galaxies. We also discuss the impact of numerical resolution on our results. Finally, in Section 5, we present a summary of our main conclusions.

\section{Simulations}
\subsection{\textsc{c-eagle} simulations}

Here we briefly summarize the main characteristics of the \textsc{c-eagle} simulations. For a more detailed description of the simulations and their subgrid model, see \citet{Barnes17} and \citet{Bahe17}.
The \textsc{c-eagle} project comprises a suite of 30 cosmological hydrodynamical zoom-in simulations of  massive galaxy clusters, spanning a mass range between $14 \leq \log_{10}M_{200}^{z=0}$[M$_{\odot}$]$\leq$ 15.4. The clusters were selected from a parent N-body low-resolution simulation with a box of size (3.2 Gpc)$^3$, first presented in \citet{Barnes17a}.
Each zoom-in simulation was performed adopting the same flat $\Lambda$CDM cosmology that was used in the \textsc{eagle} simulation \citep{Schaye15,Crain15} corresponding to  $\Omega_{\Lambda}$ = 0.693, $\Omega_m$ = 0.307, $\Omega_b$ = 0.04825, $\sigma_8$ = 0.8288, Y = 0.248 and H$_{0}$ = 67.77 km s$^{-1}$ \citep{Planck14}.

All simulations were performed with the variant `AGNdT9' of the \textsc{eagle} simulations \citep[]{Schaye15}, with a dark matter and baryonic mass resolution (gas and stars) of $m_{\rm DM}$ = 9.7 $\times$ 10$^{6}$ M$_\odot$ and $m_{\rm gas}$ = 1.8 $\times$ 10$^{6}$ M$_\odot$, respectively.
The gravitational softening was set to $\epsilon$ =  2.66 comoving kpc at $z \geq 2.8$, and set to $\epsilon = 0.7$ physical kpc at $z < 2.8$.

The zoom-in simulations were run with a high-resolution region extending at least to 5$r_{200c}$, with the 24 galaxy clusters belonging to the \textsc{hydrangea} sub-sample extending their high-resolution region to at least 10$r_{200c}$. For this study, two of the \textsc{c-eagle} clusters were removed (CE-10 and CE-27). These clusters experienced a very dramatic numerical AGN outburst at high redshift that significantly and artificially affected the overall properties of the system. 

The code used to run these models is an upgraded version of the N-Body Tree-PM SPH code \textsc{Gadget 3} \citep[described in][]{Springel05}. The modifications include updates to the hydrodynamics scheme, collectively known as `\textsc{anarchy}' (see Appendix A in \citealt{Schaye15} and \citealt{Schaller15} for details) and several subgrid physics models to simulate unresolved properties.
Radiative cooling and photoheating are implemented following \cite{Wiersma09}, assuming an optically thin X-Ray/UV background \citep{Haardt01}. Star formation is implemented stochastically based on the Kennicutt-Schmidt law \citep{Kennicutt98} in pressure law form as in \citet{Schaye08}, using the metallicity-dependent density threshold as in \cite{Schaye04}. 
Each particle is assumed to be a single-age stellar population, with a Chabrier initial mass function in the range 0.1M$_\odot$-100M$_\odot$ \citep{Chabrier03}.

Stellar evolution is modelled following \cite{Wiersma09b}, and chemical enrichment is followed for the 11 elements that most contribute to radiative cooling(i.e., H, He, C, N, O, Ne, Mg, Si, S, Ca, and Fe). The thermal energy released by stellar feedback is stochastically distributed among the gas particles surrounding the event without any preferential direction  \citep{Vecchia12}.

The galaxy formation model used in the \textsc{c-eagle} project has been widely tested for the \textsc{eagle} simulations. \citet[][]{Schaye15} showed that the model is successful at reproducing  several observable quantities that were not directly calibrated in the simulations. As an example, galaxies in the \textsc{eagle} simulation have a realistic neutral gas content fraction \citet[][]{Bahe16,Crain17}, a key property when studying the evolution of the star formation in galaxies. 
Moreover, in \citet[][]{Barnes17} it is shown that the model reproduces with reasonable good agreement important properties of the intracluster medium, such as the gas fraction-total mass relation and its metallicity. The predicted X-ray and Sunyaev-Zeldovich properties of clusters show a good match with the spectroscopic temperature, X-ray luminosity, and $Y_{SZ}$-mass relation.
Regarding the stellar content in galaxies, the cluster satellite's stellar mass function in the \textsc{c-eagle} project is in excellent agreement with observations up to $z\sim1.5$ \citep[][]{Bahe17,Ahad21}. Additionally, the satellite quenched fractions are also in reasonably good agreement with observations at $z=0$ \citep[][]{Bahe17} and subhalo masses are well matched to lensing observations \citep[][]{Bahe21}.

\subsection{Halo identification}

The main outputs from each simulation correspond to 30 snapshots spaced between $z=14$ and $0$, 28 of them  equidistant in time with a $\Delta t$ = 500 Myr. Two additional snapshots at $z=0.366$ and $0.101$ were added to facilitate comparisons with \textsc{eagle}. 
All these outputs were later post-processed to identify structures in each snapshot using the \textsc{subfind} algorithm \citep{Springel01,Dolag09} in a two-step procedure, as described below.

In order to define bound halos, first, a friends-of-friends (FoF) algorithm is applied to all dark matter particles. Baryons are then assigned to the FoF (if any) associated with their nearest dark matter particle. If a FoF halo possesses fewer than 32 dark matter particles, it is considered unresolved and discarded. As a second step, \textsc{subfind} identifies any gravitationally self-bound substructures (or `subhaloes') within an FoF halo taking dark matter and baryons into consideration. These subhaloes are identified as local overdensities using a binding energy criterion. For a more detailed description of the method, we refer to \cite{Springel01} and \cite{Dolag09}.

\textsc{subfind} identifies structures at a single point in time. Since we need to follow galaxies both in space and time an additional procedure is required to create merger trees. The trees analyzed in this work were obtained with the \textsc{spiderweb} algorithm \citep[see][Appendix A]{Bahe19}. Differently from other `merger tree' algorithms, \textsc{spiderweb} was designed with the purpose to identify descendants for as long as possible in crowded environments. For that purpose, \textsc{spiderweb} consider as descendants all subhaloes that share particles between consecutive snapshots. This treatment becomes especially relevant in groups and cluster of galaxies, where galaxy-galaxy encounters become more common. 
The satellite galaxies in our sample correspond to those satellite galaxies that are inside the $r_{200}$ of the most massive structures of each zoom-in simulation.
Throughout this paper, we will define as galaxies all those subhalos that posses an stellar mass  $M_{\rm star} > 10^8$M$_\odot$, and a total mass $M_{\rm galaxy} > 10^9$M$_\odot$. 
As a result, we ensure that each of our galaxies is resolved with at least,  1000 dark matter particles at any snapshot. 
Additionally, as we also want to understand the reasons behind the satellite quenching in our simulations, we choose to keep the mass threshold in which the star formation evolution, stellar mass function, and mass-size relation are reasonably well-calibrated for the galaxy formation model \citep[][]{Schaye15,Furlong15,Furlong17}.
To avoid detection of galaxies with no primordial origin, we only followed the evolutionary history of those galaxies whose merger trees can be continuously followed at least up to $z=2$.
It is worth noting that, as discussed by \citet{Bahe17}, the \textsc{c-eagle} simulation shows an excess of quenched galaxies in the low-mass end ($M_{\star} < 10^{9.5}$M$_\odot$). This result is believed to be a combination of environmental quenching effects and limitations related to the resolution of the simulation \citep[][]{Schaye15}. Note, however, that in our work we focus on galaxies that undergo their quenching process purely due to environmental mechanisms. This means that most of our selected galaxies would have reached a higher stellar mass at $z=0$ if not affected by the physical processes explored in this article. As a result, our conclusions are not significantly affected by the stellar mass threshold chosen to select satellite galaxies. This is further  discussed in Appendix \ref{ap:stellar_threshold}, where we explored the effects of choosing  a higher stellar mass threshold ($M_{\star} > 10^{9.5}$M$_\odot$) for the satellite selection. 
We discuss this overquenching in more detail in Section \ref{sec:overquenching} and the impact that could have on our results, in order to avoid spurious results, we will neglect the population of low-mass galaxies that reach their quenching state as centrals, and only take into account those galaxies that reach their quenched state as satellites. Nonetheless, regardless of the cluster mass, galaxies that reach their quenching state while centrals correspond to less than 10\% of all the galaxies in our cluster sample.

\begin{table}
\begin{tabular}{|lllll|}
\hline
\multicolumn{1}{|l|}{ID cluster} & \multicolumn{1}{l|}{$M_{500}^{z=0.1}$} & \multicolumn{1}{l|}{$R_{500}^{z=0.1}$} & \multicolumn{1}{l|}{$E_{\rm kin}$/$E_{\rm thermal}$} & Virialization state \\ \hline
CE-0                             & 13.905                                 & 0.65                                   & 0.11                                                 & Unrelaxed           \\
CE-1                             & 13.982                                 & 0.69                                   & 0.24                                                 & Unrelaxed           \\
CE-2                             & 13.920                                 & 0.66                                   & 0.05                                                 & Relaxed             \\
CE-3                             & 13.926                                 & 0.66                                   & 0.06                                                 & Relaxed             \\
CE-4                             & 13.853                                 & 0.62                                   & 0.11                                                 & Unrelaxed           \\
CE-5                             & 13.892                                 & 0.64                                   & 0.09                                                 & Relaxed             \\
CE-6                             & 14.128                                 & 0.77                                   & 0.09                                                 & Relaxed             \\
CE-7                             & 14.176                                 & 0.80                                   & 0.09                                                 & Relaxed             \\
CE-8                             & 14.070                                 & 0.74                                   & 0.09                                                 & Relaxed             \\
CE-9                             & 14.244                                 & 0.84                                   & 0.16                                                 & Unrelaxed           \\
CE-11                            & 14.290                                 & 0.87                                   & 0.21                                                 & Unrelaxed           \\
CE-12                            & 14.422                                 & 0.96                                   & 0.08                                                 & Relaxed             \\
CE-13                            & 14.341                                 & 0.91                                   & 0.06                                                 & Relaxed             \\
CE-14                            & 14.563                                 & 1.08                                   & 0.31                                                 & Unrelaxed           \\
CE-15                            & 14.407                                 & 0.95                                   & 0.22                                                 & Unrelaxed           \\
CE-16                            & 14.311                                 & 0.89                                   & 0.12                                                 & Unrelaxed           \\
CE-17                            & 14.537                                 & 1.05                                   & 0.25                                                 & Unrelaxed           \\
CE-18                            & 14.639                                 & 1.14                                   & 0.09                                                 & Relaxed             \\
CE-19                            & 14.586                                 & 1.09                                   & 0.30                                                 & Unrelaxed           \\
CE-20                            & 14.482                                 & 1.01                                   & 0.12                                                 & Unrelaxed           \\
CE-21                            & 14.800                                 & 1.29                                   & 0.29                                                 & Unrelaxed           \\
CE-22                            & 14.837                                 & 1.33                                   & 0.12                                                 & Unrelaxed           \\
CE-23                            & 14.426                                 & 0.97                                   & 0.26                                                 & Unrelaxed           \\
CE-24                            & 14.821                                 & 1.31                                   & 0.13                                                 & Unrelaxed           \\
CE-25                            & 15.045                                 & 1.56                                   & 0.31                                                 & Unrelaxed           \\
CE-26                            & 14.899                                 & 1.39                                   & 0.08                                                 & Relaxed             \\
CE-28                            & 14.902                                 & 1.39                                   & 0.13                                                 & Unrelaxed           \\
CE-29                            & 15.077                                 & 1.60                                   & 0.30                                                 & Unrelaxed           \\ \hline
\end{tabular}

\caption{Main properties for the 30 clusters composing the \textsc{cluster-eagle} suite. In this table are specified the $M_{500}$, $R_{500}$, and virialization state of each cluster. The $E_{\rm kin}$/$E_{\rm thermal}$ correspond the ration between the kinetic and thermal energy of the gas particles inside one $R_{500}$ as detailed in \citet{Barnes17}. Clusters with $E_{\rm kin}$/$E_{\rm thermal} <0.1$ are defined as relaxed. }
\label{tab:Cluster_Properties}
\end{table}

\subsection{Ram pressure and restoring force models}
\label{sec:RP model}
To measure the instantaneous ram pressure experienced by cluster satellites galaxies, we follow the methodology described in \citet[][]{Vega-Martinez21}. They introduce a general analytic profile for the instantaneous ram pressure experienced by satellite galaxies that is a function of their host mass and redshift. The profile is described by a damped power law as

\begin{equation}
P_{\rm ram}(M,z) = P_0(z) \left[ \frac{1}{\xi (z)} \left( \frac{r}{R_{200} }  \right)    \right]^{-\frac{3}{2}\alpha(M_{200},z )} 
\end{equation}

This depends directly on $r/r_{200}$,  
the relative distance of the satellite galaxy from the halo centre.
$P_0(z)$ sets the normalization  
of the profile; 
$\xi(z)$ determines the radial scaling, 
and the exponent $\alpha(M,z)$ regulates the dependence on 
the host halo mass according to

\begin{equation}
\alpha(M,z) = \alpha_\text{M}(z) \log\left(M_{200} h^{-1}\left[\text{M}_{\odot} \right] \right) -5.5.  
\end{equation}

Thereby, the shape of the profile is fully described by 
$P_0(z)$, $\xi(z)$, and $\alpha_\text{M}(z)$. 
Their dependence on the redshift is expressed in terms of 
the scale factor $a=1/(1+z)$, as follows:
\begin{equation}
    \log \left(\frac{P_0(z)}{10^{-12} h^2 \text{dyn\;cm}^{-2}} \right) = 7.01 a^{-0.122} - 9.1,
\end{equation}
\begin{equation}
    \xi(z) = -3.4 a^{-0.42} + 10.2,
\end{equation}

\begin{equation}
    \alpha_\text{M}(z) = 3.3\times 10^{-3} a^{1.33} + 0.512.
\end{equation}

The numerical values in these relations were obtained from a $\chi^{2}$ minimization to fit ram pressure measurements from hydrodynamical resimulations of groups and clusters of galaxies. 

Given the low scatter found in the relation by the authors, and given the great agreement between the values found with this method and direct measurement inside the clusters in our $z=0$ sample, throughout this work, we will use this expression to estimate the ram pressure acting on our satellite galaxies at any time. 

A galaxy will suffer ram pressure stripping once the ram pressure overcomes the restoring force exerted by the galaxy itself. To measure the restoring pressure, we follow the methodology presented in \citet{Simpson18}. The restoring force per area on the satellite's gas can be expressed by 
\begin{equation}
    P_{\rm rest} = \left|\frac{\partial \Phi}{\partial z_h} \right|_{\rm max} \Sigma_{\rm gas},
\end{equation}

where $\Phi$ corresponds to the gravitational potential,
$z_h$ corresponds to the direction motion of the gas (and in the opposite direction of the gas displacement),
$ \left|{\partial \Phi}/{\partial z_h} \right|_{\rm max}$ is the maximum of the derivative of $\Phi$ along $z_h$, and $\Sigma_{\rm gas}$ is the surface gas density of the satellite \citep{Roediger05}.
We adopt a simple estimate for both $\left|{\partial \Phi}/{\partial z_h} \right|_{\rm max}$ and $\Sigma_{\rm gas}$ that can be applied to all the galaxies in our sample regardless of their morphology.
Additionally, we will measure $\left|{\partial \Phi}/{\partial z_h} \right|_{\rm max}$ as
\begin{equation}
   \left|{\partial \Phi}/{\partial z_h} \right|_{\rm max} =  \mathit{v}^2_{\rm max}/\mathit{r}_{\rm max}
\end{equation}
where $\mathit{v}_{\rm max}$ is the maximum circular velocity of the galaxy and $\mathit{r}_{\rm max}$ is the distance to the center of the galaxy where the maximum circular velocity is reached. For the gas surface density, we will measure it using 
\begin{equation}
\Sigma_{\rm{gas}} = M_{\rm gas}/\pi (2r_{1/2}^{\rm gas} )^{2},
\end{equation}

where $M_{\rm gas}$ is the total gas mass of the galaxy and $(r_{1/2}^{\rm gas} )$ is the 3D radius enclosing  half of the gas mass.

As shown by \citet[][]{Vega-Martinez21}, the profile presented has the strength of providing accurate and quick ram pressure measurements for galaxies in different environments, from the outskirts (~ 2$r_{200}$) of low-mass groups to the inner parts of massive galaxy clusters. 
Additionally, the profile works in a wide range of redshift, from $0<z<2$. This profile allows us to study the ram pressure efficiency of the different structures. 
Even though this profile does not take into account the asymmetric behaviour of the ICM, the agreement between our simulations and the profile is remarkable and allows us to characterize the pressure exerted by the different haloes in which galaxies reside throughout its life (for details, see Appendix \ref{ap:rp_vega}).

\section{Properties of clusters: Quenched population and gas density profiles}

As previously discussed, it is well established that the environment can play an important role in quenching the star formation activity of galaxies, especially within the inner and denser environments of clusters. However, how the mechanisms associated with environmental quenching lead to the final quenching state is still an open question. In this work, we want to study the relation between the local density of the ICM and the moment of quenching for the satellite population.
Several criteria have been used in the past to select quenched galaxies. Following previous studies, we use a threshold  in specific star formation rate (sSFR), defined as the instantaneous SFR divided by their total stellar mass ($M_\star$) \citep[]{Weinmann10,deLucia12,Wetzel12,Wetzel13} such that if a galaxy satisfies (i) that its sSFR $\leq 10^{-11}$yr$^{-1}$ and (ii) it never reaches a sSFR greater than that threshold value again, it is classified as quenched. 
It is worth mentioning that we tried a different threshold sensitive to redshift to split our sample between star-forming and quenched galaxies, following \citet{Dave19}. Using this criterion, a galaxy is classified as quenched if its sSFR $\leq 10^{-10.8 + 0.3z}$yr$^{-1}$. We found none or little differences in our results. In order to make our results directly comparable with previous works, we choose to keep the fixed sSFR threshold.

\begin{figure*}
\centering
\includegraphics[width=\textwidth]{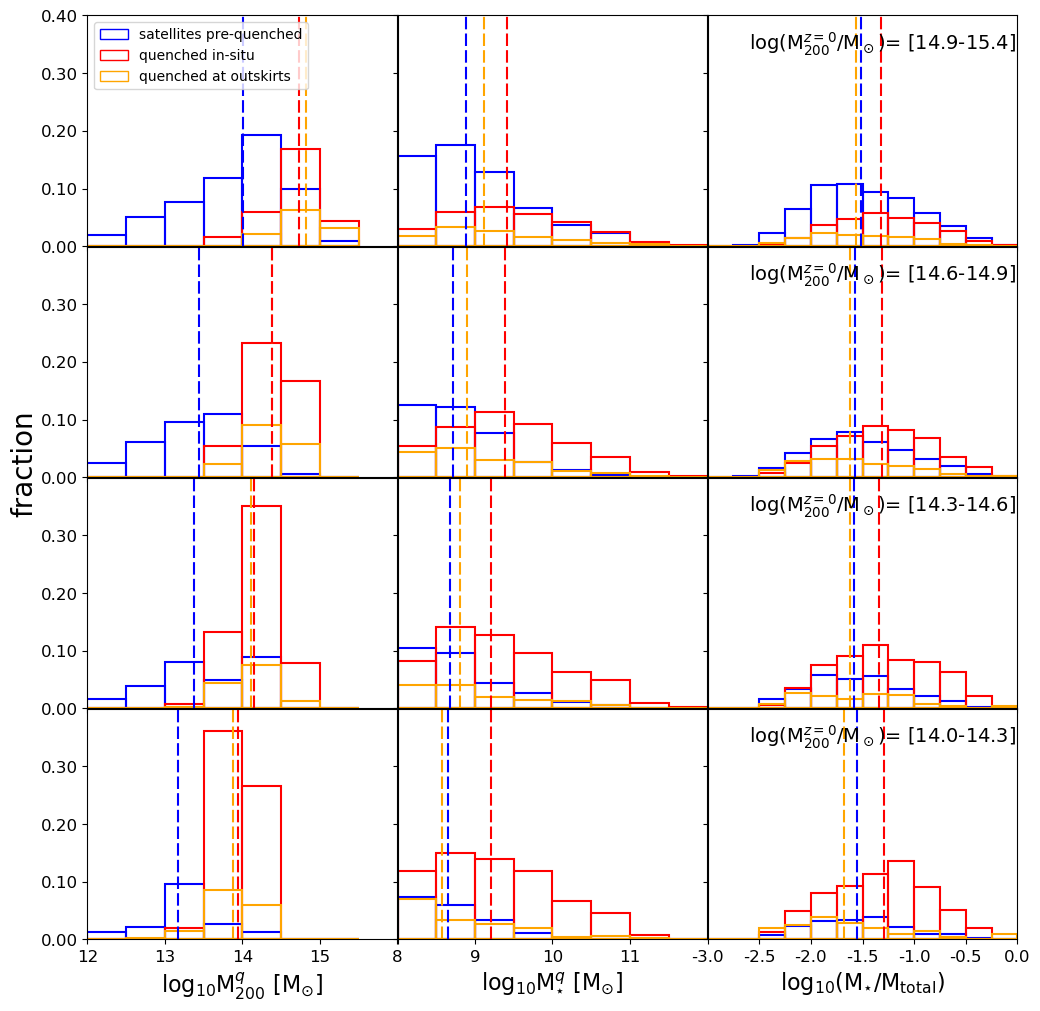}
\caption{Mass distribution of galaxies and their hosts at the moment when they quench. Each row corresponds to the results obtained after stacking the distribution of galaxies associated with clusters within (from upper to lower row) high, intermediate-high, intermediate-low and low mass bin respectively. The left-hand column shows the mass distribution of the
host of each galaxy at the moment of quenching. The middle column shows the distribution of galaxies' stellar mass at quenching time and the right-hand column shows the distribution of stellar mass fractions. The galaxies are separated into `pre-quenched as satellites' (blue), `quenched on cluster outskirts' (orange), and galaxies `quenched in situ' (red). We can see that there is a trend between the $z=0$ cluster mass and the halo mass at the quenching time for the satellite population. Nonetheless, independent of the $M_{200}$, more than 80\% of cluster galaxies get quenched inside a halo with $M_{200} \geq 10^{13.5}M_\odot$.}

\label{fig:hist_quenching}
\end{figure*}

\citetalias{Pallero19} studied the quenching history of the satellite population of galaxy clusters in the \textsc{eagle} simulation. They used the same quenching classification criteria that we will use throughout this paper.
They found that cluster satellite galaxies reach their quenching state preferentially in dense environments such as massive galaxy groups or low-mass galaxy clusters. This holds even for galaxies that become quenched before their infall into the $z=0$ host cluster and those that suffer strong drops in their SFR (but not get completely quenched) prior to their infall to any cluster. They also found that the quenched fraction in $z=0$ cluster satellites grows significantly after the first $r_{200}$ crossing. However, this trend depends strongly on the cluster mass, with low-mass clusters showing the largest increase of the quenched fraction during satellite infall. Here, we aim to determine the physical mechanisms behind satellite quenching during the $r_{200}$ crossing and to understand why low-mass clusters seem to play a larger role in this phenomenon.

\subsection{Quenching of the star formation}

Because of their high resolution and wide range of halo masses, spanning from groups to clusters, the \textsc{c-eagle} re-simulations are an excellent laboratory to understand the different environmental processes that affect the quenching of star formation and its dependence on cluster mass. Under the assumption that more massive clusters are associated with denser environments, we first search for correlations between the properties of galaxies at their quenching time and the mass of the halo where they reach their quenching state.

Following \citetalias{Pallero19}, and to improve the number statistics of our results, we stack galaxy in four bins of host cluster mass M$_{200}(z = 0)$. \citetalias{Pallero19} split the 10 most massive halos in the \textsc{eagle} simulations in three bins of mass, ranging from 14 < log$_{10}M_{200}$/M$_{\odot}$ < 14.8, and found significant differences between the properties of satellites residing in the lowest and highest cluster masses at quenching.

Taking advantage of the larger \textsc{c-eagle} sample, we will be able to see if the trends found in \citetalias{Pallero19} prevail at higher masses. To this end, we split our $z=0$ cluster sample as follows:

\begin{itemize}
    \item high mass: $14.9 < \log_{10}M_{200}/$M$_{\odot}  < 15.4$ 
    \item intermediate-high mass: $14.6 < \log_{10}M_{200}/$M$_\odot < 14.9$
    \item intermediate-low mass: $14.3 < \log_{10}M_{200}/$M$_{\odot} < 14.6$ 
    \item low mass: $14.0 < \log_{10}M_{200}/$M$_\odot < 14.3$.

\end{itemize}

The number of clusters in each mass bin from high to low mass is  8, 8, 6 and 7, respectively.

 To characterize the relation between the environment in which galaxies reside and the quenching of their star formation, we will further split our galaxy sample as follows:
 \begin{itemize}
     \item pre-quenched as satellites: quenched as satellites of another structure, i.e., outside the cluster.
     \item quenched on cluster outskirts: satellites quenched outside clusters $r_{200}$ but while part of the cluster FoF.
     \item quenched in situ: satellites quenched inside the $r_{200}$ of the final cluster.
 \end{itemize}

We note that the term pre-processing is commonly used in the literature to refer to all environmental effects that could affect galaxies prior to their accretion into clusters \citep[e.g.][]{Boselli05,Wetzel12,deLucia12,Einasto20}. This is, regardless of whether the result is the cease of their star formation or not. In this work, we are  focusing on galaxies that fully cease their star  formation activity before being accreted onto clusters. For this particular ``pre-processed" population we adopted the term ``pre-quenched''.  This definition was taken from  \citet[][]{Pallero19}, in which ``pre-processing'' accounted for those moments when galaxies experienced strong drops in their star formation rate, while ``pre-quenching'' accounted for those galaxies that cease their star formation prior to their accretion into the studied cluster.

Figure \ref{fig:hist_quenching}  shows the distribution of $z=0$ satellite properties at their quenching time, $t_{\rm q}$, for the four cluster mass bins. From the top to bottom panel, we show the results for \textit{high, intermediate-high, intermediate-low,} and \textit{low}  cluster mass bins respectively. In each panel, blue, orange, and red histograms represent the population of satellites pre-quenched as satellites, quenched on cluster outskirts, and quenched in situ, respectively. Each histogram is normalized by the total number of satellite galaxies per cluster mass bin. The dashed lines correspond to the median of each population.

 In the left-hand column, we show the distribution of satellite host mass at their $t_{\rm q}$. We can see that, regardless of the mass of the cluster, quenching tends to happen in high-mass haloes.
 We find that the typical host mass at the time of quenching increases with $z=0$ cluster mass; this change is more appreciable in the pre-quenched population. However, note that the bins in these histograms are 0.5 dex wide; i.e. the peak of the distribution is in-between 13 $\leq \log_{10}M_{200}/$M$_{\odot}  \leq 14.5$, and correspond to the typical mass range for low-mass clusters. We will further explore this in detail in Section \ref{section:withredshift}. 
 From the pre-quenched satellite population, we see that the peak in the host mass distribution at $t_{\rm q}$ moves from galaxy groups at the low-mass bin to low-mass clusters at the high-mass bin. This is a consequence of the hierarchical nature in which the clusters in our simulations are built. 
 This result also shows that galaxies that arrive already quenched to the present-day clusters reached their quenched state earlier when they entered another massive halo  (log$_{10}M_{200}/$M$_{\odot} \geq$ 12.0).

 The middle and right-hand columns show the total stellar mass ($M_\star$) and the stellar mass fraction for the galaxies at their $t_{\rm q}$. The stellar mass fraction is defined as the stellar mass divided by their total mass, $M_\star$/$M_{\rm galaxy}$. Here, the total mass includes the stellar, gas, black-holes and dark matter mass.  Galaxies quenched in situ show a higher stellar mass and stellar ratio compared to the pre-quenched and quenched at the outskirts populations. Moreover, galaxies that quenched on the outskirts of their $z=0$ clusters follow the same distribution as galaxies quenched as satellites of other clusters than their $z=0$ host, in both their stellar mass and stellar mass fraction. This suggests that the mechanisms affecting these two populations are similar. 
On the right-hand column, we can also see that pre-quenched galaxies tend to be more dark matter-dominated than their `in situ' quenched counterparts. We can also see that there is no relation between the median of the stellar to total mass ratio and the cluster mass in which galaxies reside at $z=0$, for all populations.

As previously discussed, more than 60\% of the galaxies quenched in our sample reach this state in groups or low-mass clusters; i.e. with
$13.5 \leq \log_{10}$ $M_{200}$ [M$_{\odot}$]$\leq 14.5$.
This suggests that satellites get quenched after their first interaction with the ICM of structures within this mass range, independently of any subsequent mergers into more massive clusters.
As massive clusters grow by the accretion of lower mass structures, most galaxies arrive in the final cluster already quenched in less massive progenitors.
In the following sections, we will study and discuss the possible physical processes that can lead to the quenching of galaxies in these structures. 

\begin{figure}
\centering
\includegraphics[width=0.5\textwidth]{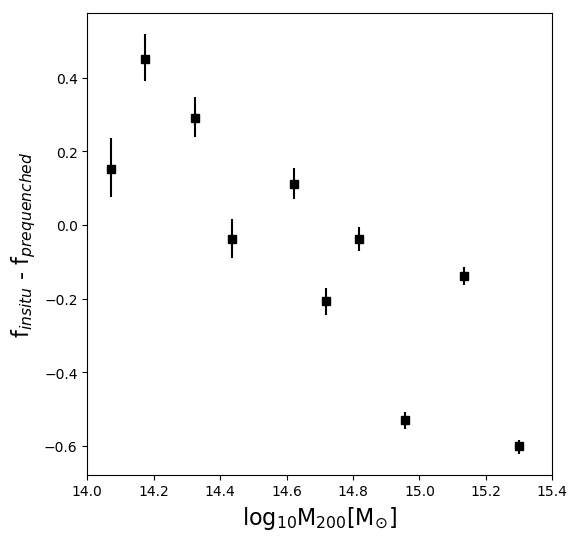}

\caption{Variation of the fraction between the population pre-quenched as satellites and the population of galaxies quenched inside the cluster's $r_{200}$ (f$_{\rm insitu}$ - f$_{\rm pre-quenched}$) as a function of their halo cluster mass. 
Each dot represents a triplet of clusters ranked by mass. The halo mass plotted corresponds to the average of each group of three clusters, with the exception of the most massive bin, in which only two clusters were used. The error bar corresponds to the binomial error associated with each measurement. 
In this plot we can see that the predominant population changes as the cluster mass grows. This result supports the scenario where low-mass clusters are  responsible for the high pre-quenching fraction found in high-mass clusters.}

\label{fig:delta_frac}
\end{figure}

Figure~\ref{fig:delta_frac} quantifies the relative abundances of the pre-quenched and in situ quenched satellite populations. To increase the satellite number statistics, and to obtain more reliable trends, we bin our cluster sample in 10 sets based on their $M_{200c}$ at $z=0$.  We divide the $30$ available clusters into ten sets of three, except for the most massive set that contains only two clusters. The figure shows the difference between the fraction of in situ quenched and pre-quenched satellites,
\begin{equation*}
\frac{N_{q}
^{\rm insitu} - N_{q}^{\rm preq}}{N_{\rm total}} = f_{\rm insitu} - f_{\rm preq},    
\end{equation*}
as a function of the mean $M_{200c}$ of each bin. 
The error bar on each dot corresponds to the binomial error associated with each measurement.
This figure clearly shows that for low-mass clusters, galaxies quenched in situ are the predominant population. On the other hand, for high-mass clusters, the population of pre-quenched satellites becomes highly dominant. At M$_{200}^{z=0} \sim$ 10$^{14.6}$M$_{\odot}$ the dominant population changes from in situ-quenched dominated to pre-quenched dominated. It should be noted that more than 60\% of cluster galaxies that are quenched at present-day reached this state in structures with 13.5 $\leq \log_{10}$M$_{200}$[M$_{\odot}$]$\leq 14.6$.

\begin{figure*}
\centering
\includegraphics[width=\textwidth]{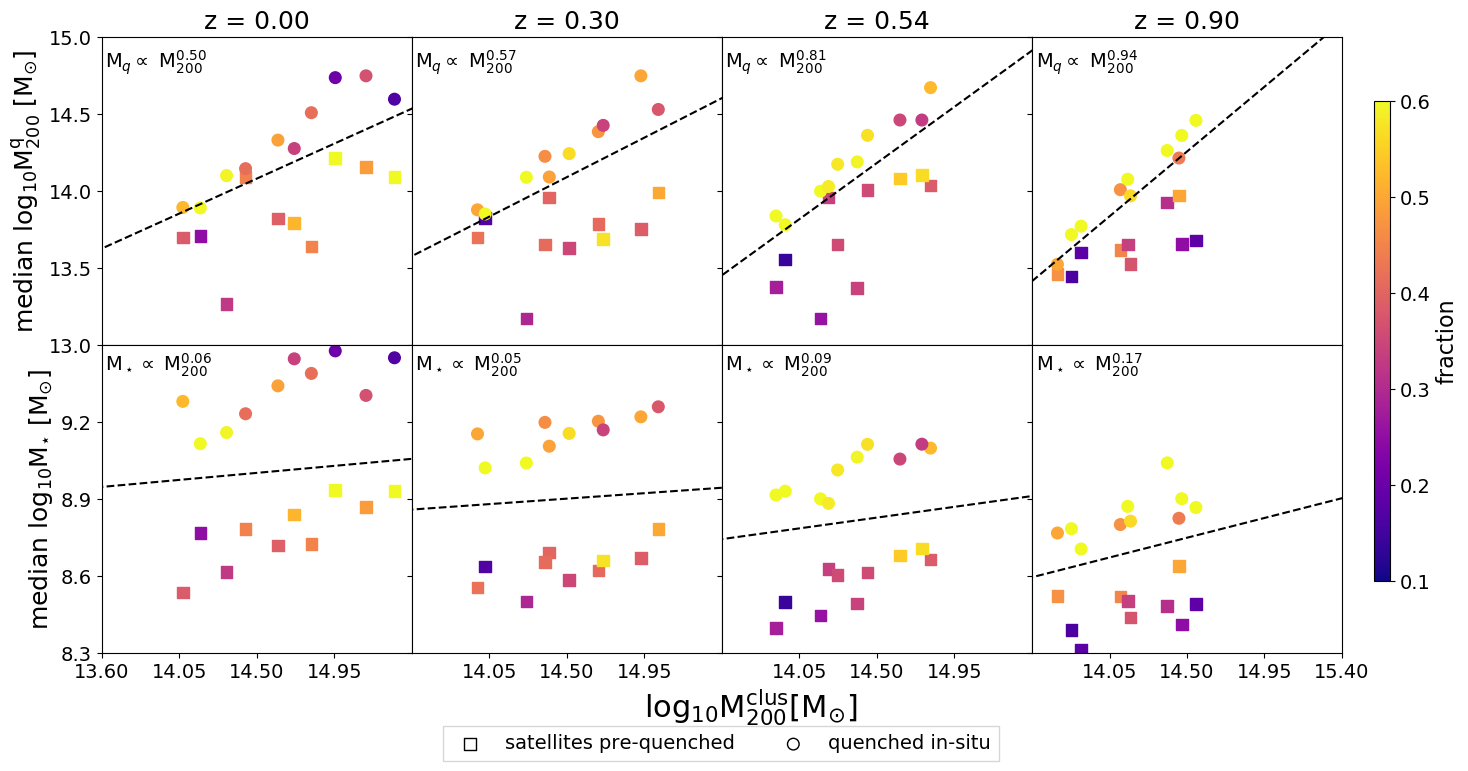}
\caption{ Median of the halo and stellar mass for the satellite galaxies at their quenching time as a function of their halo mass at z = 0.0; 0.30; 0.54; and 0.90 (from the left to right-hand panel). Squares represent the median of the pre-quenched population while the circles represent the population quenched in situ. Each dot represents a triplet of clusters ranked by mass, color-coded by the fraction that the pre-quenched or in situ quenched population represents in a given bin of mass and redshift from the whole population for those clusters. Additionally, a fit to the whole population is plotted in dashed black lines. As we look at higher redshift clusters, in situ quenching becomes dominant, and the median of the halo mass where galaxies are quenched is similar to the mass of the cluster itself. As we go down in redshift pre-quenching becomes dominant. This highlights the fact that galaxies reach their quenching state in the first transition from central to satellite galaxy. Regarding the stellar mass, as we look at galaxies at higher redshift, the stellar content is reduced systematically, while the slope of the fit shows a mild systematic growth.}

\label{fig:median_z}
\end{figure*}

\subsection{Clusters at different times}
\label{section:withredshift}

In this section, we explore the evolution of the cluster satellite properties over a range of redshifts $0<z<1$, typical for current photometric surveys.
As in the previous sections, we study the satellite population of the most massive structures in each zoom-in simulation, at each redshift.
It should be noted that their corresponding satellite galaxies can be different from the  $z = 0$ population, as many of them will be disrupted before the present-day \citep[but see ][]{Bahe19}, while others will be accreted after the selected redshift.
We choose to compare the satellite population of clusters at four different times: $z = 0.90, 0.54, 0.30$ and $0$. These correspond to lookback times of 7.5, 5.5, 3.5, and 0.0 Gyr, respectively. As before, for each satellite galaxy, we record the values of their total stellar mass and the host halo mass at their corresponding $t_{\rm q}$.

The top panels of Figure \ref{fig:median_z} show the relation between the median host mass when satellites quench and the mass of the cluster they belong to at the corresponding $z$. The different columns show the results at the four redshifts considered. Each symbol represents the median of the distribution, $\bar{M}_{200}^{q}$; squares and circles represent the population of pre-quenched and in situ quenched satellites, respectively. As before, clusters have been combined into 10 mass bins. Thus, each symbol represents the population of satellites of three clusters (two in the last bin). The colour coding indicates the fraction that each population represents with respect to the total, at the given redshift. For clarity, galaxies quenched as centrals are not shown in this figure. Nevertheless, they are considered when estimating the percentage that each population represents. 

In the top left-hand panel, we show the relation  between $\bar{M}_{200}^{q}$ and the cluster's $M_{200}$ at $z=0$. The panel shows a mild growth of $\bar{M}_{200}^{q}$ with  $z=0$ cluster mass. The dashed lines show the result of a linear regression fit, obtained combining all satellites inside each cluster-mass bin, regardless of whether they were quenched in situ or pre-quenched. Not surprisingly, this relation is stronger for the in situ quenched than for the pre-quenched satellites. From the colour coding, we can also see that, at $z=0$, the dominant population changes with cluster mass (see also Fig.~\ref{fig:delta_frac}). For low $z=0$  cluster masses, most satellites were quenched in situ,  while for their high-mass counterparts satellites arrive into the cluster pre-quenched.  However, the values of $\bar{M}_{200}^{q}$ are all within the mass range  of low-mass clusters, i.e. $\log_{10}M_{200}$~[M$_{\odot}$]$~\sim 14.0$.

As we move towards a higher redshift, we can see that the slope of the fit increases, reaching a value $\approx 1$ at $z = 0.9$. In section \ref{section:rpsection}, we will study in detail the mechanisms that produce this trend. Note that the most massive structures at this $z$ have $M_{200} \sim 10^{14.6}\rm M_\odot$. Interestingly, according to Fig.~\ref{fig:delta_frac}, this mass corresponds to the transition where the pre-quenched population at $z=0$ becomes dominant over the in situ quenched satellites. 
A comparison between the left-hand and right-hand top panels ($z=0$ and $0.9$, respectively) shows that, even at this transition mass, the in situ quenched population is more dominant at high than low redshifts. This could be produced by the different assembly histories that clusters at $z=0.90$ and $0$ have experienced. High-redshift clusters have mainly accreted star-forming galaxies residing in low-mass structures, while low-redshift clusters were able to grow by the accretion of more massive, pre-quenched substructures.

\begin{figure*}
\centering
\includegraphics[width=\textwidth]{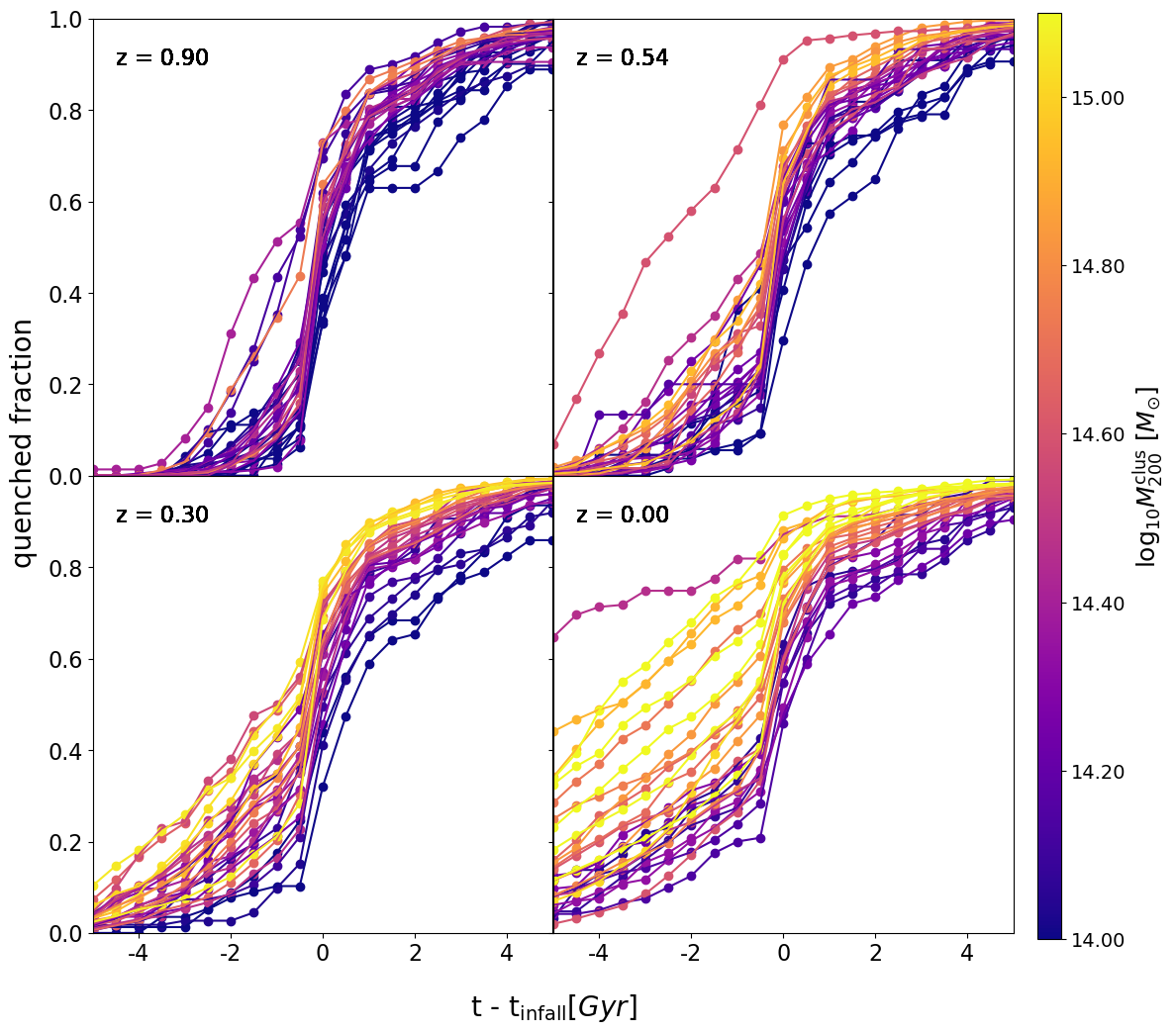}
\caption{Growth of the quenched fraction of the satellite population of galaxy clusters as a function of the normalized time-scale t - t$_{\rm infall}$  for clusters at z = 0.00; 0.30; 0.54; and 0.90. Each line represents one cluster in the simulation, color-coded by their $M_{200}$ at the given redshift. Negative (positive) values correspond to times before (after) the first crossing of the R$_{200}$ of the final cluster. 
As we look at clusters at higher redshift, we can see that the fraction of pre-quenching is reduced significantly (t-t$_{\rm infall} < 0.$). At the time of their infall (t - t$_{\rm infall}$ = 0) we can see a sharp increase in quenching and an abrupt change in the slope of the fraction of quenched galaxies. This result is even clearer when looking at higher redshifts, where even the most massive clusters present a rise in the quenched fraction $\gtrsim 60\%$ when $-1 \lesssim$ t - t$_{\rm infall} \lesssim 1 $. As we decrease the redshift, the fraction of pre-quenching increases, especially for the most massive clusters, which at $z=0.30$ presents $>50\%$ of their galaxies already quenched before accretion. And by $z=0$, clusters with $M_{200} \geq 10^{15}$M$\odot$ possess a pre-quenching fraction$>80\%$. }

\label{fig:tinf_z}
\end{figure*}

In the bottom panels, we show the median of the galaxies' stellar mass distribution, $\bar{M}_*$, at their quenching time. Here we find a clear correlation (but a very mild one) between  $\bar{M}_*$ and the cluster mass. Instead, we find a correlation between  $\bar{M}_*$ and redshift, independently of the cluster mass. 
Similar to what was found in the previous section, a difference of 0.5dex in $\bar{M}_*$ between the pre-quenched and the in situ quenched population is found at all $z$.

In order to highlight the role played by the environment in the quenching of the star formation activity of satellites, in Figure~\ref{fig:tinf_z} we show, for each cluster, the satellite quenched fraction as a function of satellite accretion time onto the corresponding cluster. The panels show the results obtained at different $z$.  Each line is associated with one of the $30$  simulated clusters, and the colour coding indicates the cluster  $M_{200}$ at the given $z$. Following P19, to generate this figure we computed, for each galaxy within the cluster $R_{200}$ (at the corresponding $z$), the time ($t_{\rm infall}$) when it first crossed the cluster's $R_{200}$. We then define, for each galaxy,  the variable $t - t_{\rm infall}$, and identify the moment when it became quenched on this new time-scale. Finally, we compute the cumulative quenched galaxy fraction as a function of $t - t_{\rm infall}$. 

The bottom right-hand panel of Fig~\ref{fig:tinf_z} shows the result obtained at $z=0$. For all clusters, we find that the satellite quenched fraction rises rapidly after infall ($t - t_{\rm infall} > 0$), regardless of the cluster mass. The jump is, however, significantly more abrupt for low-mass clusters. High-mass clusters show a larger fraction of quenched satellites for $t - t_{\rm infall} < 0$ and rapidly reach a fraction of $90\%$  after satellite infall. In general, we find that the fraction of pre-quenching varies between 20-80\%, with a strong dependence on the cluster mass. Almost every satellite galaxy reaches its quenched state 4 Gyr after their accretion time (see also \citetalias{Pallero19}). 
As previously discussed, most $z=0$ satellites get quenched while inhabiting low-mass clusters and arrive at the final massive structure without any star formation activity. As a result, the $\rho_{\rm ICM}$ of these $z=0$ massive clusters ($M_{200} \gtrsim 10^{14.6}$ M$_{\odot}$) are not playing a significant role in the overall quenching of (massive) cluster galaxies in the local Universe. Instead, their main role is to end the star formation activity of that small amount of galaxies that is not accreted in a subgroup into the cluster. 

As we move towards a higher redshift, the fraction of satellites arriving into the clusters as star-forming galaxies increases. Note that at $z=0.9$ (top left-hand panel) the typical pre-quenched satellite fraction is $\lesssim 20\%$, and that just after $t_{\rm infall}$ these fractions quickly jump to values of $\sim 70\%$. Most cluster satellites are suffering strong environmental effects for the first time in structures that are within the mass range  $ 10^{13.5} \lesssim M_{200} \lesssim 10^{14.5}$ M${_\odot}$, i.e. low-mass clusters (see also Fig.~\ref{fig:hist_quenching}) 

It should be noted that there is a clear outlier at $z=0$, that can also be well distinguished at $z=0.54$. This line corresponds to the cluster CE-9 which underwent a major merger at $z\sim 0.6$. This structure collides with another cluster of similar mass which already had a large fraction of quenched galaxies. The result is the accretion onto the main cluster of a large number of pre-quenched satellite in a very short time-scale. As a result, this cluster posses $\sim 40\%$ larger pre-quenched fraction with respect to other clusters with similar mass.

\begin{figure*}
\centering
\includegraphics[width=\textwidth]{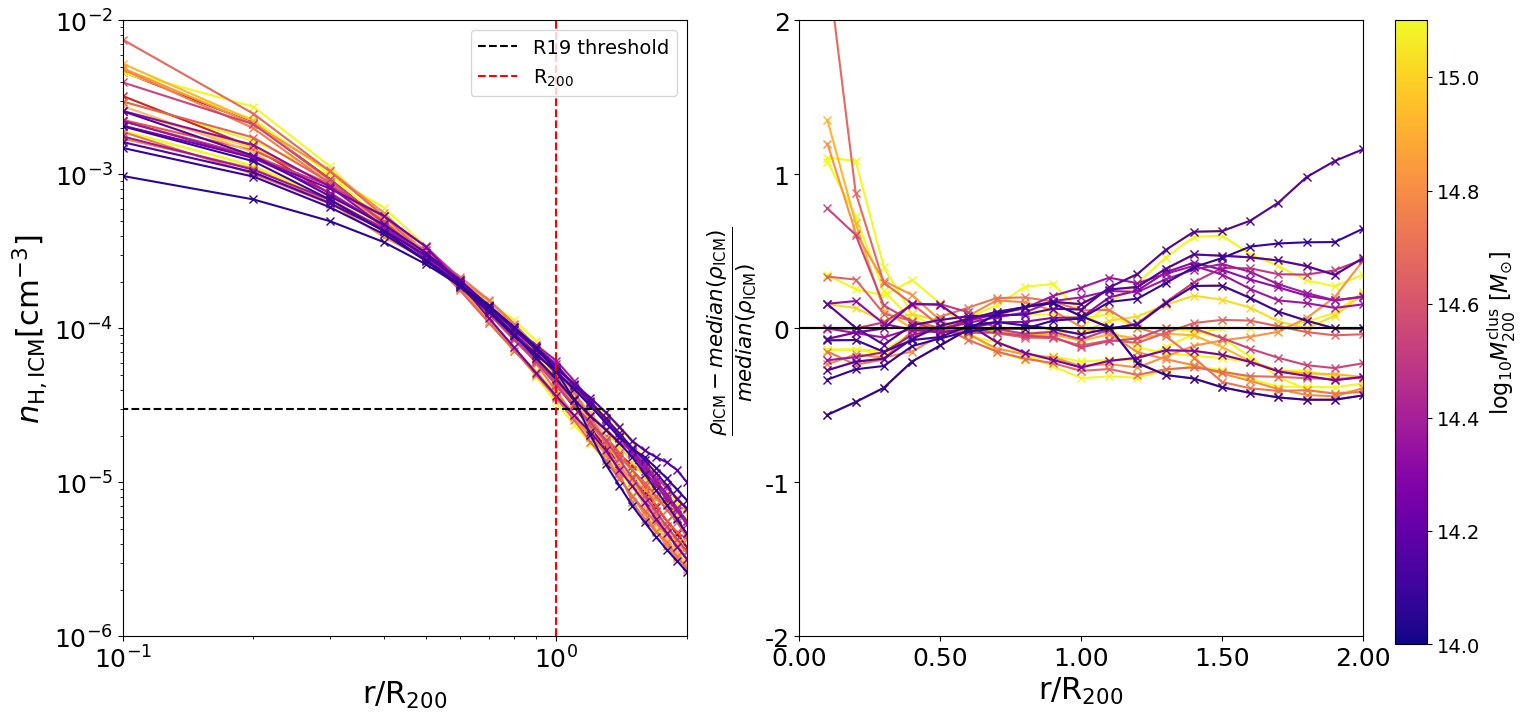}

\caption{ \textit{Left-hand panel:} azimuthally averaged cluster gas density profiles for the $z=0$ sample.
\textit{Right-hand panel:} fractional deviation from the median cluster gas density profiles.
Each line is colour-coded by cluster $M_{200}$. 
 We can see that low-mass clusters possess a more extended gas component in comparison with high-mass clusters that are more concentrated. This result is more evident when looking at the normalized distribution in the right panel.
Regardless of the cluster mass, the threshold in $\rho_{\rm ICM}$ found by \citet{Roberts19} is reached near the $r_{200}$ of our clusters.}

\label{fig:hot_cold_gas}
\end{figure*}

\begin{figure}
\centering
\includegraphics[width=0.5\textwidth]{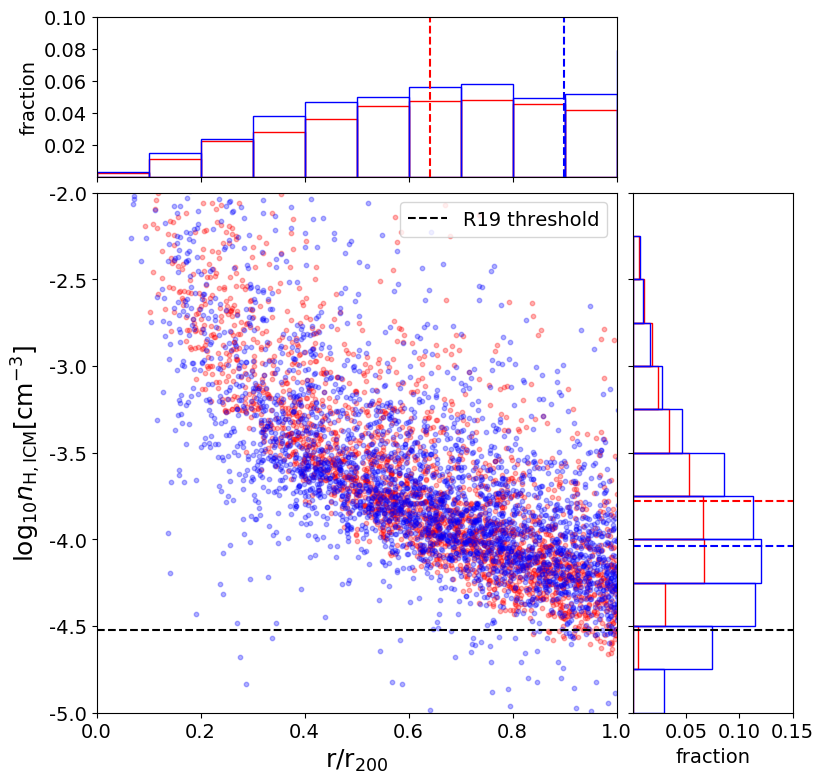}

\caption{Density of the local environment in which  galaxies reside at their quenching time as a function of the distance from the halo center in units of $r_{200}$. Blue dots correspond to  galaxies that are quenched in a different halo than the final cluster (pre-quenched) and red dots correspond to galaxies quenched in the final cluster (quenched in situ). The histograms show the distribution of density and distance separately. 
The black dashed lines show the threshold found in \citetalias{Roberts19} and the blue and red dashed lines in the histograms correspond to the median of each distribution. We can see that the majority of the population in our sample, regardless of their status as pre-quenched or in situ quenched, suffer quenching in high-density environments (log$_{10}n_{\rm H, ICM} \geq 3\times10^{-5}$[cm$^{-3}$]) with the majority of the population reaching this state in the outskirts of clusters (r$>$0.5$r_{200}$). }

\label{fig:2d_hist_rho}
\end{figure}

\begin{figure*}
\centering
\includegraphics[width=\textwidth]{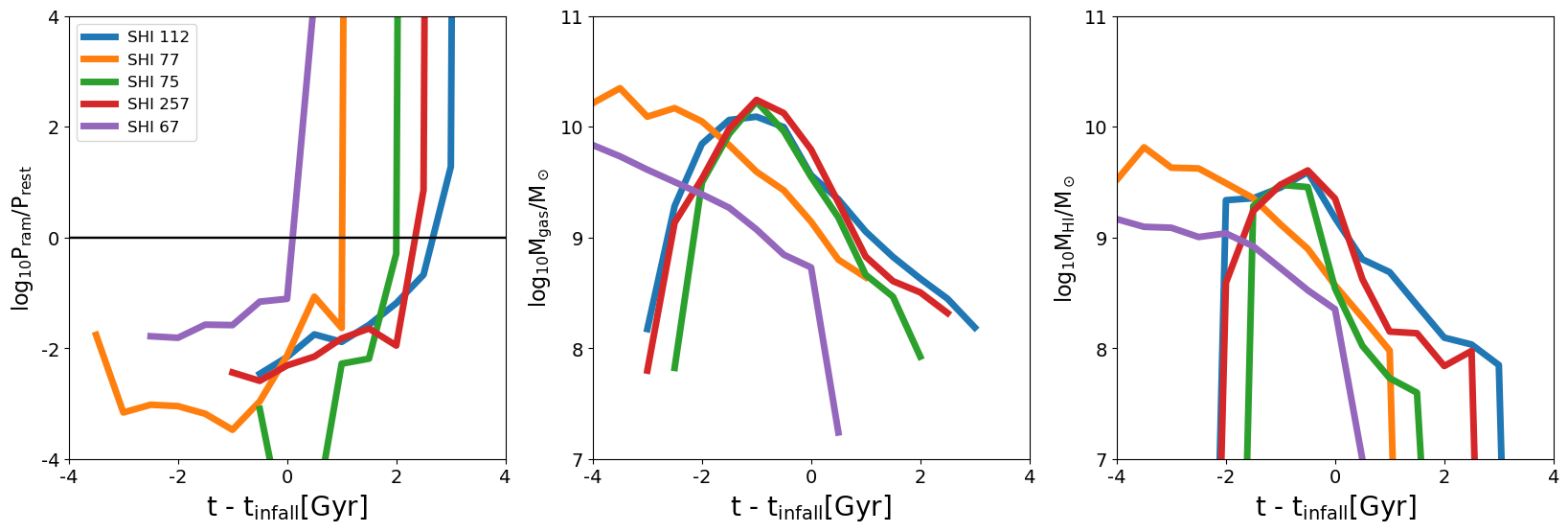}

\caption{ Evolution of properties for a subset of galaxies in our sample. SHI correspond to the SubHaloIndex associated with each galaxy in the zoom-in simulation CE-0. \textit{Left-hand panel:} ratio between ram pressure and the restoring force per area, \textit{Middle panel:} evolution of the total gas mass and \textit{Right-hand panel:} HI content in the galaxy, as a function of their time since accretion. The restoring force per area dominates over the ram pressure until they arrive to the cluster. At the time that galaxies become ram-pressure dominated, they lose most of their total gas content.  }

\label{fig:RP_examples}
\end{figure*}

\subsection{Gas density of the ICM}

Our analysis points towards a preferential cluster mass range where environmental quenching is more efficient. Interestingly, using a sample of cluster galaxies selected from the Sloan Digital Sky Survey with high-quality Chandra X-ray data, \citetalias{Roberts19} found a threshold in $\rho_{\rm ICM}$ for quenching, likely related to the effects of ram pressure stripping. Their study shows that the fraction of quenched galaxies in dense environments grows with $\rho_{\rm ICM}$. For galaxies with a stellar-mass log$_{10}$M$_{\star}$[M$_{\odot}$] > 9.9 (intermediate and high-mass galaxies), the relation between the quenched fraction and $\rho_{\rm ICM}$ can be described with a single power law. However, for galaxies in the mass range of 9.0 < log$_{10}M_{\star}$[M$_{\odot}$] < 9.9 (low-mass galaxies) a broken power law is needed. 
The `knee' in this double power law is located at densities $\rho_{\rm ICM}$ = 10$^{-28.3}$ gr cm$^{-3}$ ($n_{\rm H} = 3\times10^{-5}$[cm$^{-3}$]). According to the model presented in \citetalias{Roberts19}, when a galaxy reaches this density threshold it starts to experience a `rapid-quenching' mode. It is worth noting that, as discussed by \citet{Simpson18}, the ICM of hosts with total masses in the range 12 < log$_{10}M_{200}$[M$_{\odot}$] < 13 are inefficient at quenching this type of satellites. Thus, they are bound to reach the final quenching state within the environments of low-mass clusters and massive groups.  

To explore this scenario with our models,  in Figure \ref{fig:hot_cold_gas} we show azimuthally averaged gas density profiles of the 28 $z=0$ main clusters in the \textsc{c-eagle} simulation. Each line on the left-hand panel corresponds to the gas density distribution of one cluster in our simulation, colour-coded by their $M_{200}$ at $z=0$. The cluster-centric distances have been normalized by the corresponding $r_{200}$. Note that low-mass clusters typically have relatively more extended gas distributions than their more massive counterparts.
These differences are better highlighted in the right-hand panel, where we show
\begin{equation*}
    \Delta \rho_{\rm ICM} = \dfrac{\rho_{\rm ICM} - \bar{\rho}_{\rm ICM}}{\bar{\rho}_{\rm ICM}}. 
\end{equation*}
Here, $\bar{\rho}_{\rm ICM}$ represents, at each radius, the median of the distribution of  $\rho_{\rm ICM}$ for all clusters. These panels clearly show that the distribution of gas in high-mass clusters is more concentrated. These clusters also possess lower amounts of cold gas than lower mass objects.

The black dashed line on the left-hand panel indicates the $\rho_{\rm ICM}$ threshold introduced by \citetalias{Roberts19} and the red dashed line shows the density of each cluster at their $r_{200}$. In general, we find that low-mass clusters reach this threshold a bit further out, $\sim 1.3~R_{200}$, than high-mass clusters, at $\sim 1.0~R_{200}$. Note that, as shown in Figure~\ref{fig:tinf_z} (see also \citetalias{Pallero19}), a drastic change in the slope of the cumulative satellite quenched fraction as a function of time takes place at $(t - t_{\rm infall} = 0)$ Gyr; this is, during the first crossing of the clusters $r_{200}$.

Using the density profiles shown in Fig~\ref{fig:hot_cold_gas} we estimate, for each cluster, the local value of $\rho_{\rm ICM}$ at the time where in situ quenched satellites ceased their star formation activity. These values are obtained by interpolating the density profiles shown in Fig~\ref{fig:hot_cold_gas} at the locations where satellites are identified as quenched. 
We find that the peak of the local density experienced by each galaxy at their quenching time is  $n_{\rm H,ICM} \sim 1.8\times 10^{-4}$gr cm$^{-3}$ briefly after 0.7R$_{200}$ which roughly corresponds to  $r_{500}$ \citep{Ettori09}.

To extend these results to the pre-quenched population, we estimate the local densities where galaxies reach their quenching state, either in the final host or in an accreted substructure. This time we use the ram pressure profile shown in Section \ref{sec:RP model}. 
As discussed,  ram pressure can be written as 
$$\rm P_{\rm ram} = \rho_{\rm ICM}\varv^{2},$$
where $\rho_{\rm ICM}$ corresponds to the density of the local environment of the galaxy, and $\varv^{2}$ corresponds to the square of the relative velocity between the galaxy and its environment. Based on this, we estimate the local environment density using the analytic ram pressure profile and the velocity of the galaxies directly measured from the simulations.
The velocity is measured as the relative velocity between the galaxy and the centre of the host in which each galaxy resides at their $t_{\rm q}$.
In Fig~\ref{fig:2d_hist_rho}, we show the values of the local density at which galaxies reach their quenching state as a function of the normalized distance to the host-centre at their $t_q$. The red dots correspond to the in situ quenched population, while blue dots stand for the pre-quenched population. The histograms show both populations separately, and the red and blue dashed lines correspond to the median of each distribution. The dashed black line corresponds to the density threshold proposed by \citetalias{Roberts19}. We can see that in general, regardless of their in situ or pre-quenched condition, galaxies reach their quenching state in a local environment denser than the threshold proposed by \citetalias{Roberts19}. We can also see that the majority of galaxies reach their quenched state at the outskirts of their host ($\sim r_{200}$), and just a small percentage of the whole sample the galaxies reach their quenched state in the inner parts of their hosts ($r \lesssim 0.5 r_{200}$). It is worth to emphasize that we define the accretion time of a galaxy as the first time when a galaxy crosses the virial radii of its host cluster. This is done by individually tracing the histories of each galaxy throughout the simulation. On this way, we are sure that we are looking at galaxies that are undergoing their first pericentral passage.
We note that, given the time resolution of the simulation, we are not able to recover the exact moment and place where the galaxy gets quenched. Furthermore, for this particular result, we are not taking into account the time spent within the host in which they eventually get quenched. Nonetheless, as we see in Figure \ref{fig:tinf_z}, galaxies get quenched shortly after their first accretion event \citep[but see][for observational evidence to the contrary]{Oman21}.

\section{Discussion}

\subsection{Ram pressure as the main culprit}
\label{section:rpsection}

The \citetalias{Roberts19} `slow-then-rapid' quenching scenario proposes that galaxies first experience a slow quenching phase as they approach the cluster due to what is known as starvation \citep{Larson80}. During this phase galaxies are not able to replenish their gas reservoir and, as a result, slowly deplete their gas available to form stars. In the absence of other mechanisms, starvation would slowly lead to the final quenching of galaxies, within a typical time-scale  $\gtrsim 3$ Gyr  \citep{Peng10,Peng15}. However, and in particular for low-mass galaxies, once they reach the inner cluster region, where the ICM reaches $\rho_{\rm ICM} \sim 10^{-28.3}\rm gr~cm^{-3}$ ($n_{\rm H, ICM} \sim 3\times10^{-5}$[cm$^{-3}$]), the process switches to a  rapid quenching phase due to the effect of ram pressure. The ram pressure experienced by satellites in these inner regions becomes large enough to overcome their restoring force, triggering the rapid depletion of their gas component.
The typical quenching time-scale in this phase is of the order of $\lesssim$ 1 Gyr. As we have just shown in the previous section, this density is reached in these simulations at the outskirts of low-mass clusters ($\sim$R$_{200}$).

\begin{figure*}
\centering
\includegraphics[width=\textwidth]{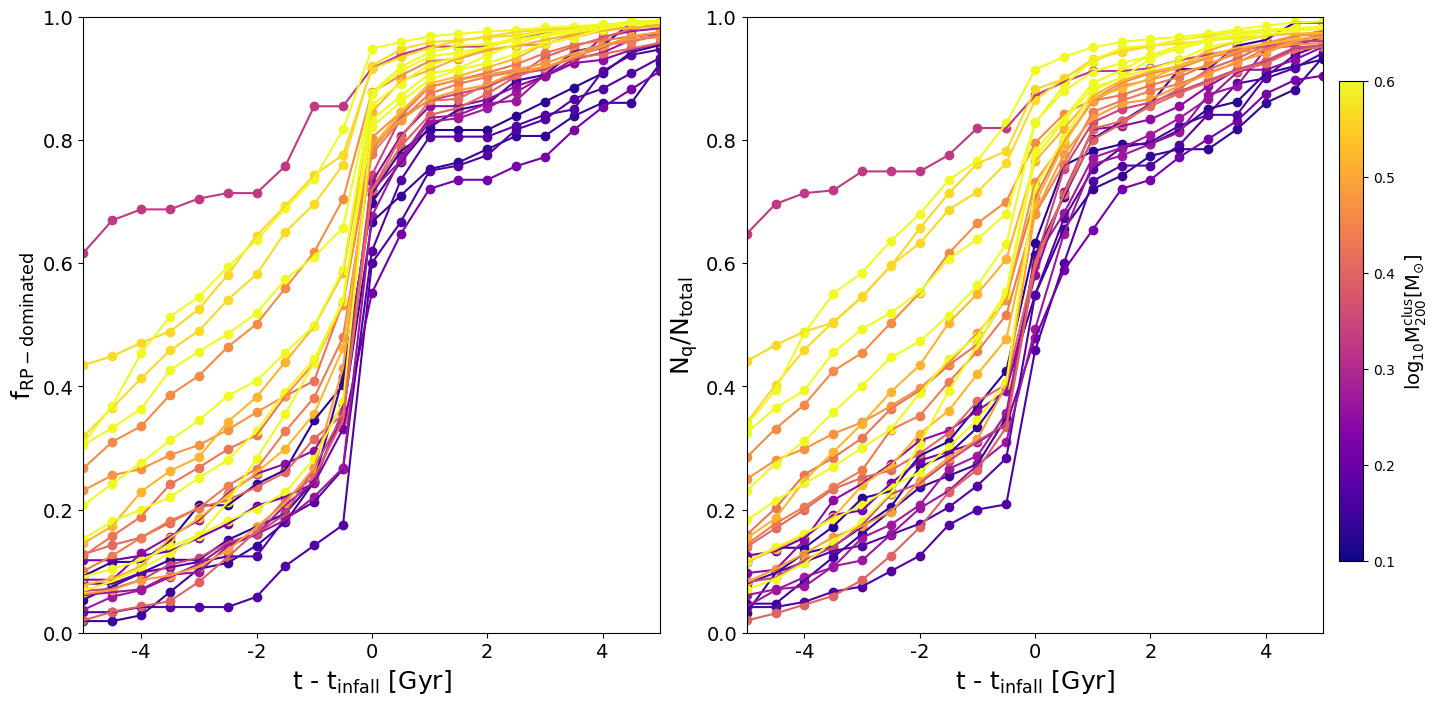}
\caption{ \textit{Left-hand panel:} evolution of the ram-pressure dominated population of galaxies as a function of time since infall.
\textit{Right-hand panel:} growth of the quenched fraction of the satellite population in clusters as a function of the time since infall.
Each line represents one cluster in the simulation, color-coded by their $M_{200}$ at z=0. These results are coincident with the evolution of the ram pressure dominated population; as galaxies become more ram pressure dominated the quenched fraction also increases. This suggests that most galaxies quenched in our sample were quenched due to ram pressure inside clusters of galaxies.}

\label{fig:tinf_q}
\end{figure*}

In Figure \ref{fig:RP_examples}, we further explore whether the rapid quenching phase we observe in our simulations is associated with the \citetalias{Roberts19} ICM density threshold, $n_{\rm H, ICM} \sim 3\times10^{-5}$[cm$^{-3}$]. In the left-hand panel of this figure, we show the time evolution of the balance between the ram pressure and the restoring force per area felt by the satellites' gas component. As an example, we focus on a subset of randomly selected galaxies. As before, the time-scale is measured as $(t - t_{\rm infall})$. This panel shows that, after crossing the virial radius of the cluster, the influence of ram pressure rapidly grows until it completely dominates over the restoring force. The middle and right-hand panels show the time evolution of the total gas and HI  content, respectively. We can see that the gas content of these galaxies starts to slowly decrease prior to accretion (slow phase) and then rapidly drops after infall (rapid phase). 
To generalize these results, we show in the left-hand panel of Figure \ref{fig:tinf_q} the cumulative fraction of galaxies dominated by ram pressure ($P_{\rm ram} > P_{\rm rest}$) for each cluster, as a function of their infall time. Note that ram pressure is estimated for each satellite based on the density of their environment at the corresponding time, using the methodology shown in Section \ref{sec:RP model}. In other words, if galaxies are arriving into the cluster as satellites of another substructure, the ICM considered is the one associated with the corresponding substructure.
For comparison, in the right-hand panel of this figure, we also show the evolution of the quenched galaxy fraction. Each line is  color-coded by the cluster $M_{200}$  at z=0. 
As we saw in the previous section, present-day satellites with masses $M_\star \gtrsim 10^9$ M$_{\odot}$ (see Fig.~\ref{fig:hist_quenching}), in low-mass clusters, tend to arrive as star-forming objects but the quenched fraction grows rapidly after infall (right-hand panel). This trend becomes progressively less pronounced as the mass of the cluster increases. The left-hand panel of this figure shows a very similar trend when considering the fraction of satellites that are ram pressure dominated (with respect to their restoring force). We can clearly see that, for low-mass clusters, the vast majority of satellites reach the cluster's $r_{200}$ for the first time, dominated by their restoring force, but this abruptly changes after infall. Instead, in more massive clusters, a significant fraction of satellites are arriving already dominated by ram pressure (and already quenched). As previously discussed, and shown in \citetalias{Pallero19}, this is due to the accretion of larger substructures into more massive clusters, bringing a significant fraction of the present-day quenched satellite population. Nonetheless, and regardless of the mass of the cluster, we see immediately after the $t_{\rm infall}$  a sharp rise in the fraction of ram pressure dominated fraction ($\sim 40\%$ in the most extreme cases). As the cluster mass grows, the rise becomes less pronounced.

\subsection{Over-quenching in \textsc{c-eagle}}
\label{sec:overquenching}
It was reported in \cite{Bahe17} that  the fraction of low-mass (log$_{10}$M$_\star$[M$_\odot$] $\leq$ 9.5) quenched galaxies in the \textsc{c-eagle} simulations is higher than observed in data. In fact, while observations show that the fraction of quenched galaxies decreases steadily as we look at lower stellar masses, in \textsc{c-eagle} this fraction rises towards low masses (See figure 6 of \citealt{Bahe17}).

In a similar way, the \textsc{eagle} simulations also posses higher fraction of quenched galaxies towards lower masses (log$_{10}$M$_\star$[M$_\odot$] $\lesssim$ 9.5), mainly due  to numerical resolution artifacts \citep{Schaye15}.
Since \textsc{c-eagle} uses the same numerical code and resolution as \textsc{eagle}, \cite{Bahe17} argue that the excess of passive galaxies found in these simulations is also mainly due to resolution effects, as overquenching is measured only in low-mass satellite galaxies.
They propose that this resolution issue can be connected to the formation of large holes ($\sim20$kpc) in the atomic hydrogen discs of many dwarf galaxies, similar to what was found in the \textsc{eagle} simulation \citep[see][]{Bahe16}. The limited resolution and high energetic feedback model implemented in \textsc{eagle} develop these large holes in the discs of dwarfs, making them more susceptible to ram pressure stripping events.

As we showed in the previous section, in general, \textsc{c-eagle} clusters possess an extended gas envelope that makes them relatively denser than the observational clusters shown in \citetalias{Roberts19}, reaching the density needed for `rapid-quenching' even at the outskirts of low-mass clusters. Given all these considerations it is straightforward to conclude that the excess of dwarf galaxies quenched inside clusters is mainly due to a combination of the extreme environment in which galaxies reside and the feedback phenomenology implemented in the code \citep[for a more detailed explication regarding the feedback model we refer the reader to][]{Schaye15,Crain15, Bahe17}. To address this issue, simulations containing galaxy clusters with higher resolution are needed. However no suitable simulation suites are available at present; \textsc{tng-50} \citep[]{TNG50-1,TNG50-2} and \textsc{romulusc} \citep{RomulusSC} only partly fulfill this requirement. These simulations lack a statistical sample of clusters to carry out our study, with just one halo with log$_{10}$M$_\star$[M$_\odot$] $\geq $ 14.
On the other hand, simulations as the ones presented in \textsc{the three hundred project} \citep[]{THREEHUNDRED}, or in the \textsc{tng-300} \citep[]{Marinacci18,Naiman18,Nelson18,Pillepich18,Springel18} can be useful to stastically study the properties of galaxies inside clusters. 
However, due to the low mass and spatial resolution, it is impossible to study in detail the mechanisms that lead to the quenching of star formation in these simulations.

\subsection{Comparison with other studies}

Previous studies based on numerical simulations have also studied the physical process intervening in the quenching of cluster satellite galaxies.  \citet[][]{Jung18} used the Yonsei Zoom-in Cluster Simulations \citep[YZiCS][]{Choi17}, a suite of 16 galaxy clusters within a mass range of $5\times10^{13} < M_{200}/M_{\odot} < 10^{15}$ ran with the adaptive mesh refinement (AMR) code \textsc{ramses} \citep[][]{Teyssier02},  to investigate the environmental mechanisms behind the origin of gas-poor galaxies inside galaxy clusters. They define their sample of  gas-poor galaxies on the $M_{gas}/M_{\star}$-$M_{\star}$ plane as the population of galaxies that deviates from the gas-rich galaxy sequence towards the low gas mass side by 2 sigmas. This sample is composed of galaxies that generally possess less than 10\% of the gas mass found on their gas-rich counterparts, at the galaxy same mass.
Similar to our results, the authors find that most galaxies get their gas component stripped within haloes with typical masses of massive groups or low-mass galaxy clusters. In particular, they split their sample between three populations: (1) galaxies that experienced their gas stripping before their accretion into the studied cluster (pre-processed population), (2) galaxies that experience a fast stripping inside the studied cluster (fast cluster processing), and (3) those galaxies that experience a slow stripping inside the final cluster (slow cluster processing). They found that 33.7\% of the galaxies in their sample suffered pre-processing. This population is dominated by satellite galaxies that underwent their processing inside haloes with  typical masses of massive groups. An additional 42.7\%  experienced ``fast cluster processing'', lossing most of their gas mass before the first pericentric passage. These galaxies usually (but not exclusively) have low gas mass at their accretion time in comparison to field galaxies, and where satellites of other haloes before  accretion. Finally, 23.7\% of the galaxies  retain their gaseous components for several Gyrs after their accretion (after their first pericentral passage). This population is mainly composed of galaxies that were central and gas-rich before being accreted into the final cluster. The authors show that galaxies can lose their gaseous component slowly if they are accreted into low-mass clusters ($\leq10^{14}M_{\odot}$) on circular orbits, i.e. that do not approach the cluster centre close enough. 
This slow cluster processing is particularly relevant for massive groups ($M_{200}< 10^{14}M_{\odot}$) and rapidly lose relevance with cluster mass. On the other hand, both pre-processing and fast cluster processing rise in importance with cluster mass, in a similar fashion to what was shown in our work.

By combining the star formation histories of observed galaxies with the evolutionary tracks of simulated galaxies, some studies have tried to constrain the accretion history of individual cluster galaxies to better understand their quenching process. Based on satellite orbital model extracted from the \textsc{hydrangea} and the YZiCS simulations \cite{Oman21} and \citet{Rhee20}, respectively, were able to characterize the different stages of the quenching process proposed by \citet[][]{Wetzel13}; i.e.  the ``delayed-then-rapid'' quenching model. The quenching time-scale obtained by both authors, for the observed sample of galaxies, is larger than what is found in our work and in \citet{Jung18}, even though the same simulation were considered. In particular, \citet[][]{Rhee20} found that the total time-scale for quenching to take place is around $\sim 5$ Gyr. In the other hand, \citet[][]{Oman21} find that the quenching timescales for satellites in clusters could be as large as $>5.5$ Gyr after their first pericentral passage. This is much larger than the time-scales observationally  measured  by other authors \citep[eg.][]{Wetzel13,Rhee20}.

Within this context, the quenching time-scale estimated from observations is generally larger than what estimated from  hydrodynamical simulations \citep[][]{Jung18,Pallero19,Lotz19}. As we show in Figure \ref{fig:tinf_z}, galaxies in our simulation reach their quenching state rapidly after their accretion into the cluster. Regardless of the cluster mass, $\geq 70\%$ of galaxies in our sample reach their quenching state within $\leq 1$ Gyr after their accretion into the cluster. 
Similarly, in \citet[][]{Lotz19} using the hydrodynamical simulations \textit{Magneticum Pathfinder}\footnote{\url{www.magneticum.org}}
the authors found that within 1Gyr of their accretion into clusters, the large majority of galaxies reach their quenching time. This usually can be related to the time when galaxies reach their first pericentral passage, and they argue that this is probably related to a ram pressure stripping event, as was probed within our work. 
Additionally, when taking into consideration more massive galaxies ($M_{\star} \geq 1.5\times10^{10}M_{\odot}$) the quenching time-scale becomes larger ($t_{q}\sim 2-3$ Gyr). This latter result is similar to the quenching time-scale that we found when using a larger threshold in stellar mass, as shown in Appendix \ref{ap:stellar_threshold}.

Is important to note however that, even though both \citet[][]{Rhee20} and \citet[][]{Oman21} consider pre-processing as an important aspect in galaxy evolution, they cannot directly account for the pre-processing effects as we do in our work. 
Given the nature of combining data from observations and orbits from simulations, in their work it is not possible to measure if the observed satellites inhabited other massive halo prior to their accretion into their final cluster. Furthermore, both \citet[][]{Rhee20} and \citet[][]{Oman21} measure the quenching time-scale as the moment when galaxies start to decrease their star formation rate, up to the moment when their star formation rate is completely depleted. Instead, in our work we consider the time-scale between the moment when galaxies crosses the virial radii of a cluster for the first time, an their last star formation rate event.
It should be noted that, as stated in \citet[][]{Pallero19}, galaxies suffer strong drops in their star formation several Gyr before their accretion into the present-day clusters. However, despite this strong drops experienced, they remain forming stars at slow pace up to their accretion into a massive halo ($\sim 10^{14}M_{\odot}$).

These discrepancies in the quenching time-scales highlights the importance to include the pre-processing in models to constrain the accretion history of galaxies. Accounting for episodes in which galaxies underwent the extreme environmental conditions provided by galaxy clusters, prior to their accretion into the present-day host, can improve the agreement between theoretical and observational works. Finally, most simulations nowadays agree that the pre-processed, or pre-quenched galaxies as referred in this work, represent the largest satellite population of massive clusters at the present-day. As examined in the IllustrisTNG simulations, the pre-processing is a key aspect needed to explain the large number of passive galaxies in clusters nowadays \citep[][]{Gu20,Donnari20}.

\section{Summary and Conclusions}

In this paper, we have studied the evolution of the quenched fraction of satellites in the galaxy clusters of the \textsc{c-eagle} simulation. \textsc{c-eagle} is a suite of $30$ cosmological hydrodynamic zoom-in simulations of massive galaxy clusters in the mass range of 14.0$\leq$ log$_{10}M_{200}/$M$_\odot$ $\leq$ 15.4 at $z=0$, using the state-of-the-art \textsc{eagle} galaxy formation code.
We select satellite galaxies with stellar mass log$_{10}M_\star/$M$_\odot$ $\geq $ 8 that at $z=0$ are bound to a main cluster and are located inside the cluster's $r_{200}$.
First, using the merger trees obtained with the \textsc{spiderweb} post-processing software (see details in \citealt{Bahe19}), we study the time evolution of the quenched population (sSFR $< 10^{-11}$yr$^{-1}$) as they traverse different environments. 
To study the quenching of star formation of galaxies inside and outside the main cluster, we define three types of populations:

\begin{itemize}
    \item pre-quenched as satellites: quenched as satellites in a structure outside the final $z=0$ cluster;
    \item quenched on cluster outskirts: quenched outside the cluster $R_{200}$ while in cluster FoF; and
    \item quenched in situ: quenched inside the main clusters $R_{200}$.
\end{itemize}

We found that regardless of the final cluster mass, most  galaxies ($>80\%$) reach their final quenching state inside dense structures (log$_{10}M_{200}/$M$_\odot$ $\geq $ 13.5); the remaining galaxies ($\lesssim 20\%$) are quenched in a medium-sized halo (12.0 $<$ log$_{10}M_{200}/$M$_\odot$ $<$ 13.5) as satellites. 
We study the relation between the median of the distribution of galaxy properties at their quenching time and the cluster final mass. 
The first thing to notice is that, regardless of the cluster mass, quenching occurs mainly in halos of log$_{10}M_{200}/$M$_\odot$ $\sim$ 14; in fact, as the cluster mass increases the predominant population in our sample changes from galaxies quenched in situ, to galaxies pre-quenched. This highlights the fact that the quenching of the star formation happens in low-mass clusters rather than in the most massive structures, even though they have the environmental conditions to strip the gas reservoirs of these satellites.  Most of the galaxies that these massive clusters accrete are already quenched due to their earlier accretion onto another low-mass cluster.

Regarding their stellar content, galaxies quenched inside the cluster contain 0.5 dex higher stellar masses than the pre-quenched population and show 0.3 dex higher stellar mass fraction.

We also studied the gas density profiles of clusters and the relation between the local density and the quenching suffered by satellite galaxies. 
Our clusters possess similar gas profiles. There are no cool-core clusters in the sample and all are widely hot-gas dominated. 
In general, high-mass clusters show a more concentrated hot gas component, while low-mass clusters exhibit a more extended envelope. In the case of low-mass clusters, they reach the critical threshold in $\rm \rho_{ICM}$, discussed by \citet{Roberts19} at $\sim$ 1.3$r_{200}$. 
This suggests that galaxies in our sample experience the 'rapid-quenching' scenario starting at the outskirts of these clusters. When comparing to the time evolution of the quenched fraction we found that, at the moment of the first $r_{200}$ crossing, the quenched fraction increases rapidly, with low-mass clusters being the ones showing the most extreme change in this fraction; for these, the rise in their quenched fraction ($\sim60\%$) is produced in a very short time-scale ($\lesssim$ 1 Gyr). After this period, the fraction  grows more slowly ($\sim 15\%$ within $1\lesssim t - t_{\rm infall} \lesssim 4$ Gyr). 
High-mass clusters contain higher fractions of quenched galaxies regardless of their accretion time. In this case, the effect of crossing the cluster's virial radius is not as significant as in the case of low-mass clusters. This is due to the earlier passage of the galaxies through the high-density structure of previous substructures. Nevertheless, we still found a measurable rise in the fraction of quenched galaxies after infall into their final massive clusters. 
1Gyr after infall, the fraction of quenched galaxies rise from 10\% to 40$\%$, and between [1-4]Gyr after their infall, the fraction of quenched galaxies grow at a much slower rate than in low-mass clusters ($\lesssim10\%$). This result is slightly different at high redshift, where most of our simulated clusters are significantly less massive. At higher $z$, we find that most galaxies reach their quenching state in situ, regardless of the cluster mass. This is a consequence of the different assembly history of clusters at high and low redshift. In other words, our results show that galaxies in the mass range of 9.0 < log$_{10}M_{\star}$[M$_{\odot}$] < 9.9 reach their quenching state inside the first dense structure they fall into. Regardless of $z$ and cluster mass, 4 Gyr after their infall, almost all galaxies ($\gtrsim 90\%$) are quenched.

Even though the excess of quenching found in our results, especially at the outskirts of clusters for low-mass galaxies, can be related to the limited resolution of the simulation,
we show that the ram pressure experienced by intermediate and high-mass galaxies at the moment when they reach this threshold in gas density is high enough to strip their gas content, shortly after their first infall.

\section*{Acknowledgements}
We want to thank the anonymous referee for the insightful comments provided, that greatly improved the quality of this manuscript.
DP acknowledges financial support through the fellowship ``Becas Doctorales Institucionales ULS", granted by the ``Vicerrector\'ia de Investigaci\'on y Postgrado de la Universidad de La Serena. DP also acknowledges partial support from Comit\'e Mixto ESO-Gobierno de Chile. DP and FAG acknowledge financial support from the Max Planck Society through a Partner Group grant. FAG and DP acknowledge financial support from CONICYT through the project FONDECYT Regular Nr. 1211370.
YMB acknowledges funding from the EU Horizon 2020 research and innovation programme under Marie Sk\l odowska- Curie grant agreement 747645 (ClusterGal) and the Netherlands Organisation for Scientific Research (NWO) through VENI grant 639.041.751. 
NP acknowlodges the support from BASAL AFB-170002 CATA, CONICYT Anillo-1477 and Fondecyt Regular 1191813. CVM acknowledges support from ANID/FONDECYT through grant 3200918.
The \textsc{c-eagle} simulations were in part performed on the German federal maximum performance computer ``Hazel-Hen'' at the maximum performance computing centre Stuttgart (HLRS), under project GCS-HYDA / ID 44067 financed through the large-scale project ``Hydrangea'' of the Gauss Center for Supercomputing. Further simulations were performed at the Max Planck Computing and Data Facility in Garching, Germany. This work used the DiRAC@Durham facility managed by the Institute for Computational Cosmology on behalf of the STFC DiRAC HPC Facility (www.dirac.ac.uk). The equipment was funded by BEIS capital funding via STFC capital grants ST/K00042X/1, ST/P002293/1, ST/R002371/1 and ST/S002502/1, Durham University and STFC operations grant ST/R000832/1. DiRAC is part of the National e-Infrastructure.

\section*{data availability}
The data presented in the figures are available upon request from
the corresponding author. The raw simulation data can be requested
from the \textsc{c-eagle} team \citep[]{Bahe17,Barnes17}





\bibliographystyle{mnras}
\bibliography{quench_final} 



\appendix

\section{A higher stellar mass threshold for galaxy selection}
\label{ap:stellar_threshold}

Here we discuss the variations that our main results have when selecting only galaxies with a stellar-mass greater $M_{\star}> 10^{9.5}$M$_\odot$.

\begin{figure}
\centering
\includegraphics[width=0.5\textwidth]{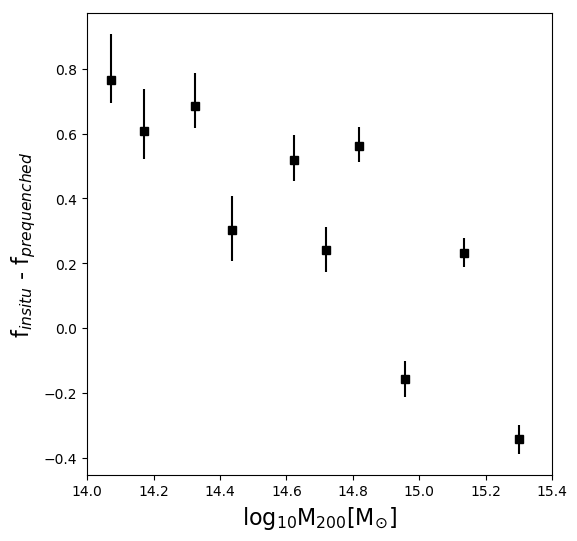}
\caption{Variation of the fraction between the population pre-quenched as satellites and the population of galaxies quenched inside the cluster's $r_{200}$ (f$_{\rm insitu}$ - f$_{\rm pre-quenched}$) as a function of their halo cluster mass, when selecting satellites with stellar mass greater than $M_{\star} \geq 10^{9.5}M_{\odot}$. 
Each dot represents a triplet of clusters ranked by mass. The halo mass plotted corresponds to the average of each group of three clusters, with the exception of the most massive bin, in which only two clusters were used. The error bar corresponds to the binomial error associated with each measurement. 
In this plot we can see that the predominant population changes as the cluster mass grows. This result supports the scenario where low-mass clusters are responsible for the high pre-quenching fraction found in high mass clusters.}

\label{fig:delta_frac_95}
\end{figure}

Figure \ref{fig:delta_frac_95} shown the variation in the predominant population between pre- and in situ-quenched as a function of the cluster mass. The error bar corresponds to the binomial error associated with each value. As shown in this figure, when selecting galaxies using a higher mass threshold the same results discussed throughout the article are found. The importance of pre-quenching grows towards higher cluster mass, becoming dominant for massive clusters. Nevertheless, it should be noted that the mass at which the predominant population changes shifts towards higher cluster masses, as expected.

\begin{figure*}
\centering
\includegraphics[width=\textwidth]{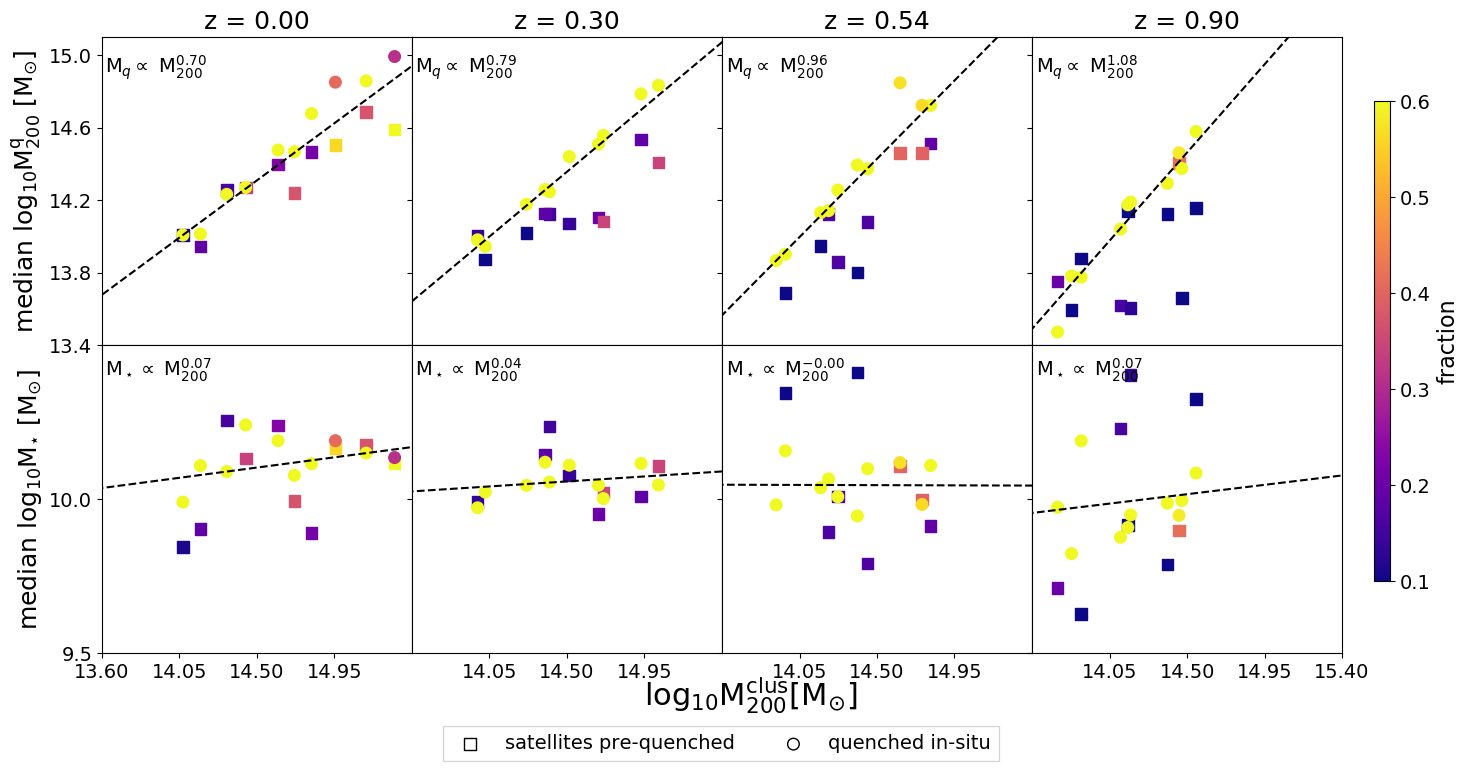}
\caption{ Median of the halo and stellar mass for the satellite galaxies at their quenching time as a function of their halo mass at z = 0.0; 0.30; 0.54; 0.90 (from left- to the right-hand panels). 
Satellites were selected using a threshold in stellar mass $M_{\star} \geq 10^{9.5}M_{\odot}$.
Squares represent the median of the pre-quenched population while the circles represent the population quenched in situ. Each dot represents a triplet of clusters ranked by mass, colour-coded by the fraction that the pre-quenched or in situ quenched population represents in a given bin of mass and redshift from the whole population for those clusters. Additionally, a fit to the whole population is plotted in dashed black lines. As we look at higher redshift clusters, in situ quenching becomes dominant, and the median of the halo mass where galaxies are quenched is similar to the mass of the cluster itself. No relation can be found between the stellar mass at quenching time and the cluster mass can be found.}

\label{fig:median_z_95}
\end{figure*}

Figure \ref{fig:median_z_95} shows the 
relation between the median host mass when satellites quench and the mass of the cluster they belong to at the corresponding $z$. The different columns show the results at the four considered redshifts. Each symbol represents the median of the distribution, $\bar{M}_{200}^{q}$; squares and circles represent the population of pre-quenched and in situ quenched satellites, respectively. As before, clusters have been combined into 10 mass bins. Thus, each symbol represents the population of satellites of 3 clusters (2 in the last bin). The colour coding indicates the fraction that each population represents with respect to the total, at the given redshift. 
The dashed lines show the result of a linear regression fit, obtained combining all satellites inside each cluster-mass bin, regardless of whether they were quenched in situ or pre-quenched.  In this case, in the top panel, as we move towards a higher redshift, the slope of the fit reach a value $\approx 1$ at $z = 0.54$, earlier than what was found when selecting galaxies with stellar mass $M_{\star}> 10^{8}$M$_\odot$. 
In the bottom panels, we show the median of the galaxies' stellar mass distribution, $\bar{M}_*$, at their quenching time. In this case, the results do not show any correlation with cluster mass.

\begin{figure*}
\centering
\includegraphics[width=\textwidth]{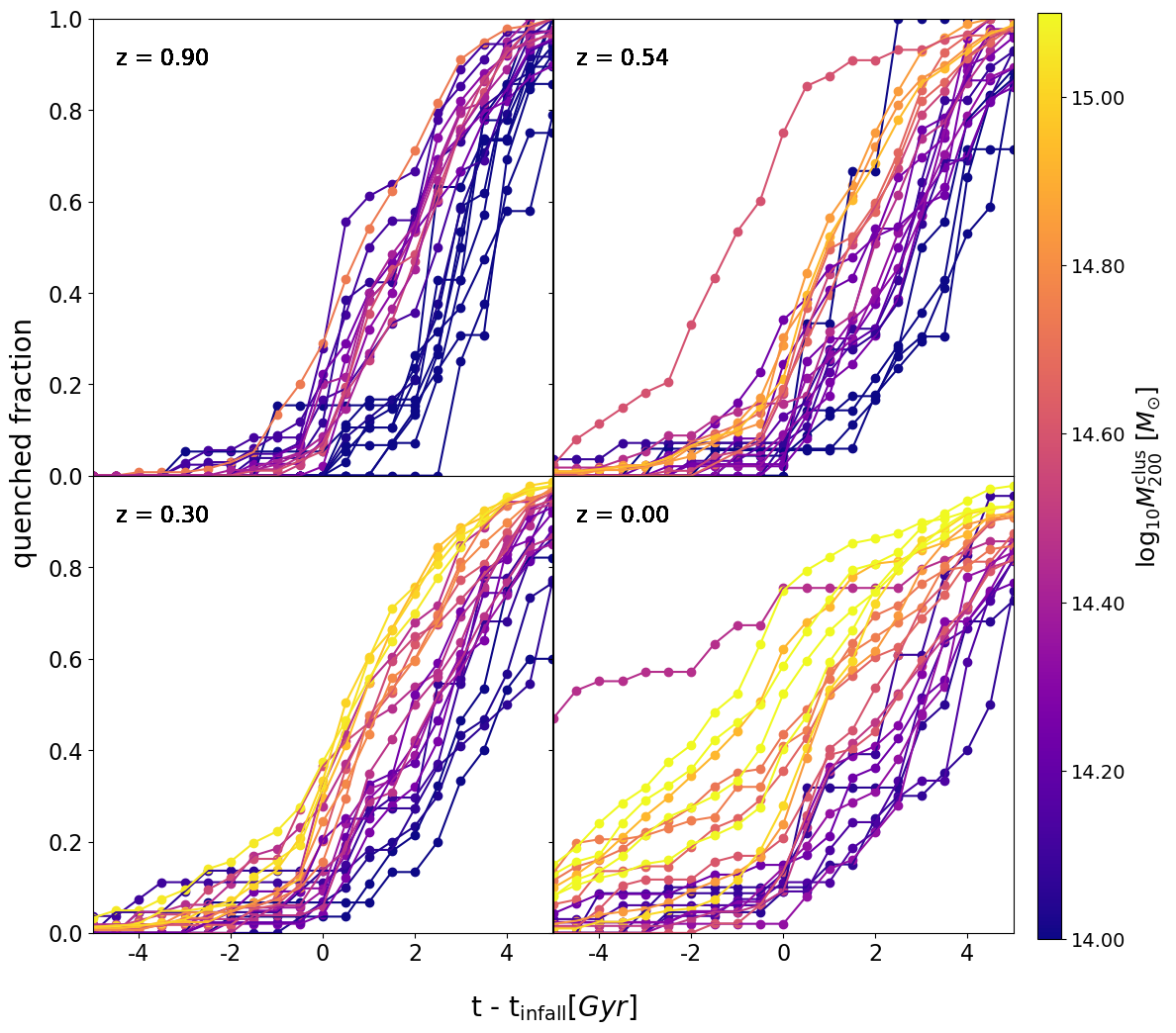}
\caption{Growth of the quenched fraction of the satellite population of galaxy clusters as a function of the normalized time scale t - t$_{\rm infall}$  for clusters at z = 0.00; 0.30; 0.54; 0.90. Each line represents one cluster in the simulation, color-coded by their $M_{200}$ at the given redshift. Negative (positive) values correspond to times before (after) the first crossing of the R$_{200}$ of the final cluster. 
Galaxies were selected using a threshold in stellar mass $M_{\star} \geq 10^{9.5}M_{\odot}$.
As we look at clusters at higher redshift, we can see that the fraction of pre-quenching is reduced significantly (t-t$_{\rm infall} < 0.$). At the time of their infall (t - t$_{\rm infall}$ = 0) we can see a sharp increase in quenching and an abrupt change in the slope of the fraction of quenched galaxies.}

\label{fig:tinf_z_95}
\end{figure*}

Figure \ref{fig:tinf_z_95}, shows the growth of the quenched fraction as a function of galaxy's accretion time, for all clusters in our sample at redshift $z = 0.00; 0.30; 0.54; 0.90$. Each line corresponds to one of our clusters and are colour-coded by their $M_{200}$ at given $z$. We can see a strong variation in the quenching fraction when galaxies cross the virial radii of a cluster for the first time, and that the fraction of pre-quenched galaxies grow with cluster mass, as we show throughout the manuscript using the lower mass threshold in stellar mass ($M_{\star} > 10^8M_{\odot}$).

\section{Agreement between the analytic profile and the simulations}
\label{ap:rp_vega}

\begin{figure}
\centering
\includegraphics[width=0.5\textwidth]{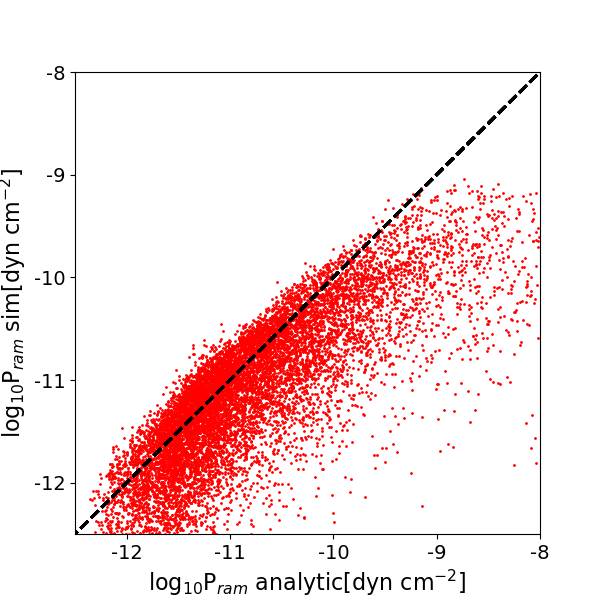}

\caption{A comparison between the ram pressure measured from the simulations using the gas density profiles of all main clusters and the relative velocity of the galaxies with respect to the BCG and the ram pressure measured with the profile provided in \citet[][]{Vega-Martinez21}. The black dashed line shows the value if the relation were one to one. As we can see both values are in great agreement, especially for the low ram pressure region.
}

\label{fig:rp_sim_vega}
\end{figure}

As mentioned in section \ref{sec:RP model}, instead of directly using the simulated gas distributions, we used an analytic profile to measure the ram pressure experience by the satellites in our sample. The implementation of this analytic method has the important advantage of significantly reducing computational costs. 
As we show in Figure \ref{fig:rp_sim_vega}, and further discussed in \citet{Vega-Martinez21},  both methods to compute ram pressure profiles show, statistically, very good agreement. The figure compares the result of measuring ram pressure profiles using an spherically averaged gas density profile extracted from the 30 \textsc{c-eagle} simulations, and that obtained from the analytic profile. Results are shown up to $R_{200}$, distance that we used in this work to define membership. 
As we can see from this Figure, 
 the ram pressure measured with the spherically averaged density profiles and the analytic formulae are in good agreement, especially for the regions where the ram pressure becomes relevant (log$_{10}$P$_{ram} < 10$[dyn cm$^{-2}$]).

\bsp	
\label{lastpage}
\end{document}